\documentclass[11pt]{article}
\usepackage[left=2.5cm,right=2.5cm,top=2.5cm,bottom=2.5cm,a4paper]{geometry} 
\usepackage{graphicx}
\usepackage{amsmath,amssymb}
\usepackage[affil-it]{authblk}
\newcommand{\gev}{\,{\rm GeV}}

\newcommand{\vevx}[1]{\left<{#1}\right>}

\usepackage{bm}
\usepackage{cite}
\usepackage{comment}

\begin{document}

\title{Neutralino dark matter in gauge mediation}

\author{Michihisa Takeuchi$^{(a)}$\thanks{takeuchi@mail.sysu.edu.cn}}
\author{Norimi Yokozaki$^{(b)}$\thanks{n.yokozaki@gmail.com}}
\author{Junhao Zhu$^{(b)}$\thanks{junhaozhu@zju.edu.cn}}

\affil{{\small (a) School of Physics and Astronomy, Sun Yat-sen University, 519082 Zhuhai, China}}

\affil{{\small (b) Zhejiang Institute of Modern Physics and Department of Physics, Zhejiang University, Hangzhou, Zhejiang 310027, China}}

\date{}

\maketitle

\begin{abstract}
\noindent
We explore the potential of neutralino dark matter within the framework of gauge-mediated supersymmetry (SUSY) breaking. Gauge mediation offers an advantage over gravity mediation, as SUSY CP/flavor problems are avoided more easily, leading to significantly milder fine-tuning for electroweak symmetry-breaking compared to gravity mediation. In our models, the lightest neutralino, as the lightest SUSY particle (LSP), is a viable dark matter candidate, assuming a gravitino mass of $\mathcal{O}(100)~\mathrm{TeV}$. The models are formulated in five-dimensional space-time, where the SUSY breaking field and the matter fields are placed on separate branes to avoid issues related to flavor and CP violation, and the radion is stabilized by a supersymmetric Goldberger--Wise mechanism. Four distinct neutralino dark matter scenarios are studied: bino-wino coannihilation, higgsino dark matter, wino dark matter, and entropy-diluted bino dark matter. For each case, we determine the allowed parameter spaces and evaluate their consistency with existing experimental limits. Additionally, we examine the potential for testing these models through future investigations at the High-Luminosity Large Hadron Collider (HL-LHC) and through dark matter direct detection experiments.
\end{abstract}

\clearpage

\section{Introduction}

The Minimal Supersymmetric Standard Model (MSSM) is one of the most compelling extensions of the Standard Model, as it stabilizes the large hierarchy between the weak scale and high-energy scales, provides a natural framework for force unification, and offers viable dark matter candidates. In particular, the stabilization of hierarchies is a crucial feature, given the presence of multiple high-energy scales such as the inflation scale, the right-handed neutrino mass scale, and the Peccei-Quinn symmetry breaking scale. Most alternative solutions for stabilizing the weak scale require treating each hierarchy among these scales separately.

The SUSY particles acquire masses through soft SUSY-breaking mass parameters. 
Since the operators associated with these mass parameters are not necessarily flavor-blind or CP-invariant, they are constrained by observations such as electric dipole moments, $K$-$\bar{K}$ mixing, and $\mu \to e \gamma$.  
In particular, $K$-$\bar{K}$ mixing imposes a stringent constraint on SUSY particles: the mass scale of the SUSY particles must exceed $\sim 1000$ TeV~\cite{Ciuchini:1998ix} if the CP and flavor violation are $\mathcal{O}(1)$.
While this still significantly alleviates the hierarchy problem compared to non-SUSY frameworks, it becomes evident that lighter SUSY scenarios, which evade the SUSY flavor and CP problems, are more favorable.

Gauge mediation~\cite{Dine:1981za, Dimopoulos:1981au, Dine:1981gu, Nappi:1982hm, Dine:1993yw, Dine:1994vc, Dine:1995ag} provides a solution to the SUSY flavor problem, as the soft SUSY breaking masses are generated through gauge interactions. It can also offer a robust solution to the SUSY CP problem: if the phases of the $B$-terms of the messenger fields and the Higgs $B$-term are aligned, the common CP phase of the soft SUSY breaking mass parameters can be eliminated by a field redefinition, thereby resolving the SUSY CP problem~\cite{Rattazzi:1996fb, Gabrielli:1997jp, Evans:2010ru, Moroi:2011fi, Evans:2022oho}. The Higgs boson mass of 125 GeV is explained with the stop mass of $\sim 10$\,TeV~\cite{Ellis:1990nz, Okada:1990vk, Haber:1990aw, Okada:1990gg}. (See also~\cite{Slavich:2020zjv}.) Since the degree of fine-tuning of the electroweak symmetry scale with 10\,TeV stop mass is much milder than the case with 1000\,TeV stop mass by a factor of $10^{-4}$, there still exists an advantage of gauge mediation models, together with the predictability of the model as the SUSY particles' masses are determined just by field theoretic perturbative calculation. Note that the absence of CP-violating phases remains crucial, as the electric dipole moment of the electron requires electroweak SUSY particles to be as heavy as $\mathcal{O}(100)$\,TeV for $\tan\beta = \mathcal{O}(10)$ (see e.g.~\cite{Moroi:2013sfa}), based on the latest experimental constraint $|d_e/e| < 4.1 \times 10^{-30}~\mathrm{cm}$~\cite{Roussy:2022cmp}.

In gauge mediation, the lightest supersymmetric particle (LSP) is typically the gravitino, whose mass is significantly smaller than the masses of the MSSM particles. In the minimal gauge mediation model, the gravitino mass is estimated as  
\begin{eqnarray*}  
	m_{3/2} \sim 16\pi^2 m_{\tilde g} \frac{M_{\rm mess}}{M_P} \approx 6.6\, {\rm keV} \left( \frac{m_{\tilde g}}{10\,{\rm TeV}} \right) \left( \frac{M_{\rm mess}}{10^7 \gev} \right),  
\end{eqnarray*}  
where $M_{\rm mess}$ is the messenger scale, $M_P$ is the reduced Planck mass, and $m_{\tilde g}$ is the gluino mass. The gravitino mass is typically less than $\sim 1$\,GeV. Unless $R$-parity is violated, the gravitino is stable and serves as a candidate for dark matter.

However, gravitinos tend to be overproduced, and to avoid this overproduction, there is a stringent upper bound on the reheating temperature $T_R$~\cite{Moroi:1993mb, Bolz:2000fu, Eberl:2020fml, Eberl:2024pxr}
\begin{eqnarray*}  
	T_R \lesssim 5\,{\rm GeV} \, \left(\frac{m_{3/2}}{10\,{\rm keV}}   \right)
	\left(\frac{10\,{\rm TeV}}{m_{\tilde g}}\right)^2.
\end{eqnarray*}  
Here, we consider the model-independent production of gravitinos through scattering processes involving particles in the thermal bath.

Motivated by the need to avoid the gravitino overproduction problem, we consider models of gauge mediation with a gravitino mass greater than 100 TeV, ensuring it is not the LSP and decays quickly into MSSM particles. This model is realized within an extra-dimensional framework. Similar to gravity mediation, the neutralino becomes the LSP and serves as a viable dark matter candidate.  
While there is a resemblance to our previous model~\cite{Asano:2015kvj}, the models proposed in this paper are more simplified and primarily focus on neutralino dark matter. (See also \cite{Nomura:2001ub} for an earlier attempt in this direction.) In contrast to these earlier studies, we give a fully explicit messenger sector, in which discrete symmetries fix the allowed interactions (Sec.~\ref{sec:model}), consistently embed a supersymmetric radion stabilization mechanism compatible with the SUSY-breaking sector (Sec.~\ref{sec:radion}), and provide a systematic, quantitative comparison of four complementary neutralino dark matter scenarios against current and projected experimental sensitivities.

We examine four scenarios:  
(i) bino-like dark matter with a nearly degenerate mass with the wino,  
(ii) higgsino-like dark matter,   
(iii) wino-like dark matter,
and 
(iv) bino-like dark matter with an entropy production mechanism.

This paper is organized as follows. Section 2 introduces a gauge mediation model within a five-dimensional framework and identifies a consistent parameter space that supports the bino-wino coannihilation scenario. Section 3 explores the prospects of higgsino dark matter and wino dark matter within the context of this model. Section 4 discusses a scenario involving bino dark matter, where entropy production dilutes the relic abundance of the bino. Section 5 presents the stabilization of the radion in this five-dimensional setup, and Section 6 is devoted to conclusions and further discussions.

\section{Model for bino-wino coannihilation} \label{sec:model}

We consider a five-dimensional (5D) $\mathcal{N}=1$ supersymmetric model compactified on an $S^1/\mathbb{Z}_2$ orbifold, described by the action:
\begin{eqnarray}
	S = \int d^4 x \int_0^L dy  \sqrt{g}
	\left[
	\mathcal{L}_0 \delta(y)
	+ \mathcal{L}_L \delta(y-L)
	+ \mathcal{L}_B
	\right],
\end{eqnarray}
where $\mathcal{L}_0$ contains the supersymmetry (SUSY) breaking field $Z$, $\mathcal{L}_L$ includes the messenger fields and the MSSM matter fields, and $\mathcal{L}_B$ comprises gauge singlet fields that mediate SUSY breaking to the messenger sector. In this setup, we focus on scenarios with a heavy gravitino mass of $\mathcal{O}(100)$ TeV. To suppress flavor violations, the MSSM fields and the SUSY breaking fields are localized on separate branes. The radion stabilization is discussed in Sec.~\ref{sec:radion}.

\paragraph{SUSY breaking}

For SUSY breaking, we consider gravitational SUSY breaking, where SUSY is broken by imposing the condition that the cosmological constant vanishes~\cite{Izawa:2010ym}. We consider the Lagrangian 
\begin{eqnarray}
	\mathcal{L}_0 \ni \int d^4 \theta \Phi^\dagger \Phi \left[-3 M_P^2 + M_*^2 g(\rho) \right] + 
	\int d^2 \theta \Phi^3 \mathcal{C} + h.c.,
	\label{eq:gravsusy}
\end{eqnarray}
where $\rho = Z + Z^\dagger$ and $\Phi = \phi(1+F_\Phi \theta^2)$ is a conformal compensator.
The four-dimensional Planck mass, $M_P \approx 2.4 \times 10^{18}\,{\rm GeV}$, is related to the five-dimensional Planck mass, $M_5$, as $M_P^2 = M_5^3 L$. The mass parameter $M_*$ is a cutoff, which is somewhat smaller than $M_P$. A slightly lower cutoff enhances the moduli mass and the $F$-term component of $Z$ compared to the gravitino mass~\cite{Asano:2015kvj, Harigaya:2017akm}.  
The cutoff $M_*$ may be close to $M_5$, $M_*/M_P \sim 1/10$ for $L^{-1}$ around $10^{15}$\,GeV.
The superpotential in Eq.~\eqref{eq:gravsusy} contains only a constant term, as we impose the shift symmetry, $Z \to Z + i \mathcal{R}$ ($\mathcal{R}$ is a real constant). With the shift symmetry, the Lagrangian relevant to SUSY breaking contains only real parameters after taking $\mathcal{C}$ to be real through a $U(1)_R$ rotation.

The $F$-terms are given by (see appendix of \cite{Harigaya:2017akm})
\begin{eqnarray}
	F_Z &=& -\frac{3 f' m_{3/2} M_P^2}{|\phi|^2 ((f')^2 - f f'')} ,
	\nonumber \\ 
	F_\Phi &=& \frac{3 f'' m_{3/2} M_P^2}{|\phi|^2 ((f')^2 - f f'')} ,
\end{eqnarray}
where $f(\rho) \equiv - 3M_P^2 + M_*^2 g(\rho)$.
By taking  
$|\phi|^2 = \left< 1-M_*^2 g/(3 M_P^2) \right>^{-1} \approx 1$ for $M_*/M_P \sim 0.1$, we go to the Einstein frame.
Requiring the vanishing cosmological constant,
\begin{eqnarray}
	V = - 3 \phi^3 F_\Phi \mathcal{C} = 0, \label{eq:fphi}
\end{eqnarray}
we obtain $f''(\rho)=g''(\rho)=0$ at the minimum. The stable minimum is obtained for $g^{(3)}(\rho)=0$ and $g^{(4)}(\rho) < 0$.
The fine-tuning of the function $g(\rho)$ can be regarded as the fine-tuning of the cosmological constant.
By using the conditions for $g(\rho)$ at the minimum, the SUSY breaking $F$-term is given by~\footnote{ For canonically normalized $Z$, $F_Z^{\rm can} \simeq -\sqrt{3} m_{3/2} M_P$. }
\begin{eqnarray}
	\vevx{F_Z} \simeq  -\frac{M_P^2}{M_*^2} \frac{3 m_{3/2}}{\vevx{g'(\rho)}},
\end{eqnarray}
which is larger than the gravitino mass scale, due to the enhancement of $M_P^2/M_*^2$. Because of $F_\Phi=0$, SUSY breaking from anomaly mediation is absent.
The scalar component of \( \rho \) has a mass of the order of \( m_{3/2} M_P^3/M_*^3 \) after canonical normalization.

Gravitational loops and loops involving bulk fields induce the violation of sequestering, and as a result, scalar masses are generated as~\cite{Gherghetta:2001sa}:
\begin{eqnarray}
	m_{\rm scalar}^2 \sim \frac{1}{16\pi^2} \frac{\Lambda_{\rm eff}^2 }{M_P^2} |F_Z|^2 \sim \frac{1}{16\pi^2} \frac{L^{-2}}{M_P^2} M_{\rm mess}^2 \, ,
	\label{eq:sequestering_violation}
\end{eqnarray}
where $\Lambda_{\rm eff}$ is the effective cut-off of the loop momentum. If this contribution is too large compared to those from gauge mediation, it becomes dangerous because these contributions can be flavor-violating. Flavor-violating soft SUSY breaking masses induce flavor-violating processes, such as $\mu \to e \gamma$. Using the latest result from MEG-II, ${\rm Br}(\mu \to e \gamma) < 4.1 \times 10^{-13}$~\cite{MEGII:2023ltw}, the off-diagonal element must be smaller than the diagonal part of the slepton mass squared by $\sim 10^{-3}$ for a slepton mass of $\sim 1$\,TeV.
Requiring that the above contributions are smaller than $\mathcal{O}(1000\gev^2)$, 
the compactification scale, $L^{-1}$, needs to be smaller than
\begin{eqnarray}
	(2\, \mathchar`- 3) \times 10^{15}\gev \left(\frac{10^6 \gev}{M_{\rm mess}}\right).
	\label{eq:L_bound}
\end{eqnarray}

Finally, we discuss the effect of gravitino decay on Big Bang Nucleosynthesis (BBN). The gravitino decays into MSSM particles with a decay rate given by (see, e.g., Ref.~\cite{Moroi:1995fs}):
\begin{eqnarray}
	\Gamma \simeq \frac{193}{384\pi} \frac{m_{3/2}^3}{M_P^2}.
\end{eqnarray}
This decay rate corresponds to a gravitino lifetime of:
\begin{eqnarray}
	\tau_{3/2} \approx 2.4 \times 10^{-2}\, {\rm s}
	\left( \frac{100\, {\rm TeV}}{m_{3/2}} \right)^3.
\end{eqnarray}
Since this lifetime is shorter than the onset of BBN, the impact of gravitino decay on BBN is negligible~\cite{Kawasaki:2008qe}. 
The thermally produced gravitinos decay to the LSPs, contributing to the LSP abundance as
\begin{eqnarray}
	\Delta \Omega_{\rm LSP} = \frac{m_{\rm LSP}}{m_{3/2}} \Omega_{3/2}.
\end{eqnarray}
Since $m_{\rm LSP}/m_{3/2} \sim 10^{-3} \mathchar`- 10^{-2}$, the LSP from gravitino decay is only a small fraction compared to the thermally produced LSPs for $T_R \lesssim 10^{8} \mathchar`- 10^{9}$\,GeV~\cite{Eberl:2020fml}.

\paragraph{Messenger sector}

\begin{eqnarray}
	\mathcal{L}_L &\ni& \int d^4 \theta \Phi^\dagger \Phi e^{-K/3M_P^2} + 
	\int d^2 \theta \Phi^3 W_{\rm mess} + h.c. \nonumber \\
	&\simeq& \int d^4 \theta e^{-K/3M_P^2} + 
	\int d^2 \theta W_{\rm mess} + h.c.,
\end{eqnarray}
where $K$ is a K\"ahler potential for the MSSM fields and the messenger field.

The messenger sector consists of brane-localized chiral superfields $\Psi_5, \bar\Psi_5$ (a ${\bf 5}$ and $\bar{\bf 5}$ of $SU(5)$), $\Psi_1, \bar\Psi_1$ (charge $\pm1$ under $U(1)_Y$, in the $SU(5)$ normalization), and $\Psi_3, \bar\Psi_3$ (a fundamental and anti-fundamental of $SU(3)_c$), together with bulk fields $X_i$ ($i=1,3,5$). These gauge quantum numbers are those of the product-group unification framework~\cite{Yanagida:1994vq, Hotta:1995cd}, based on $SU(5)_g \times SU(3)_h \times U(1)_h$, under which $U(1)_Y$ and $SU(3)_c$ are realized as diagonal subgroups of $U(1)_g\times U(1)_h$ and $SU(3)_g\times SU(3)_h$, respectively, with $\Psi_1$ and $\Psi_3$ originally charged under $U(1)_h$ and $SU(3)_h$; this is discussed further below. Since a Weyl spinor cannot be defined in five dimensions, each $X_i$ is promoted to a hypermultiplet consisting of two 4D $\mathcal{N}=1$ chiral superfields, $X_i$ and $X_i^c$. We assign $Z_2$-even parity to $X_i$ and odd parity to $X_i^c$, so that only $X_i$ has a zero mode.

In terms of these fields, the superpotential for the messenger sector is given by
\begin{eqnarray}
	W_{\rm mess} = \lambda_5 X_5 \Psi_5 \bar{\Psi}_5 + \lambda_1 X_1 \Psi_1 \bar{\Psi}_1
	+ \lambda_3 X_3	\Psi_3 \bar{\Psi}_3
	+ \sum_{i=1,3,5} \frac{\kappa_i}{3}X_i^3, \label{eq:mess1}
\end{eqnarray}
where the coupling constants, $\kappa_i$, are taken as real positive without loss of generality.
The superpotential in Eq.~\eqref{eq:mess1} is not consistent with the \( SU(5) \) grand unified theory but is compatible with the product group unification described above.

The form of Eq.~\eqref{eq:mess1} is enforced by imposing, for each $i=1,3,5$, an independent discrete symmetry $Z_{3,i}$, under which $X_i$, $\Psi_i$, and $\bar\Psi_i$ all carry charge $1$, while all other fields (including $X_j$, $\Psi_j$, $\bar\Psi_j$ for $j\neq i$, the MSSM matter fields, and $H_u$, $H_d$) are neutral. Under $Z_{3,i}$, the Yukawa coupling $X_i \Psi_i \bar\Psi_i$ (total charge $3\equiv0\pmod 3$) and the cubic term $X_i^3$ (charge $3\equiv0\pmod 3$) are invariant, while a linear term in $X_i$, cross terms such as $X_1 X_3$, and couplings to the Higgs sector such as $X_5 H_u H_d$ all carry nonzero charge under at least one $Z_{3,i}$ and are therefore forbidden. The same symmetry $Z_{3,5}$ also forbids renormalizable mixing between the messenger fields and the MSSM matter fields, such as $H_{\bf 5}\Psi_{\bar{\bf 5}}$ used in Sec.~\ref{sec:bino_entropy} to trigger a late messenger decay; there, $Z_{3,5}$ is broken by a small spurion to generate this mixing, playing the role of the messenger parity breaking discussed below Eq.~\eqref{eq:wmix}.

The Lagrangian includes the term
\begin{eqnarray}
	\mathcal{L}_0 &\ni&   \int d^4\theta  \Phi^\dagger \Phi   \sum_{i=1,3,5} h_i (\rho) |X_i|^2
\nonumber \\	 
&\simeq&  \int d^4\theta \sum_{i=1,3,5} h_i (\rho) |X_i|^2.
\end{eqnarray}
Here, we omit direct couplings between $\rho$ and $X_i^c$ since $X_i^c$ does not have a zero mode.
After taking canonical normalization of $X_i$,\footnote{
The canonically normalized field, $X'_i$, is obtained as 
$X'_i = \sqrt{1+h_i}(1 + \frac{h'_i}{1+h_i} F_Z \theta^2) X_i$
} the soft breaking mass terms for $X_i$ is obtained as
\begin{eqnarray}
	V_{\rm soft} \ni \sum_{i=1,3,5} \left[ m_{i}^2 |X_i|^2
	+ \left(A_i X_i \frac{\partial W}{\partial X_i} + h.c.\right) \right],
\end{eqnarray}
where
\begin{eqnarray}
	 	A_i \simeq \frac{h_i'}{1+h_i} F_Z, \ \ 
	 	m_i^2 \simeq A_i^2  - \frac{1}{2} \frac{h_i''}{1+h_i} |F_Z|^2. 
\end{eqnarray}
The messenger mass $\hat{M}_i=M_i + B_i M_i \theta^2$, is given by
\begin{eqnarray}
	M_i = \lambda_i \vevx{X_i}, \ 
	B_i = -\kappa_i \vevx{X_i} - A_i \,.
\end{eqnarray}
Here, $M_i \sim B_i \sim |F_Z|$. Note that, because of the $A$-terms, the angular direction of $\vevx{X_i}$ is fixed.

After integrating out the messenger fields, superpartners obtain masses. The gaugino masses are  
\begin{eqnarray}  
	M_{\tilde{b}} &\simeq& \frac{g_1^2}{16\pi^2} (B_5 + B_1)   ,
	\nonumber \\  
	M_{\tilde{w}} &\simeq& \frac{g_2^2}{16\pi^2} B_5 ,
	\nonumber \\  
	M_{\tilde{g}} &\simeq& \frac{g_3^2}{16\pi^2} (B_5 + B_3),  
\end{eqnarray}  
and the slepton masses are  
\begin{eqnarray}  
	m_{\tilde L}^2 &\simeq& \frac{1}{256 \pi^4} \left[\frac{3}{2}g_2^4 B_5^2 + \frac{3}{10} g_1^4 (B_5^2 + B_1^2) \right]  ,
	\nonumber \\  
	m_{\tilde E}^2 &\simeq& \frac{1}{256 \pi^4} \left[ \frac{6}{5} g_1^4 (B_5^2 + B_1^2) \right].  
\end{eqnarray}
The other soft masses are listed in Appendix~\ref{sec:ap_a}.
Note that the gaugino masses are all real.
This is crucial to avoid the stringent constraint of the electron electric dipole moment, which forces electroweak SUSY particles to be as heavy as $\mathcal{O}(100)$ TeV.
There is no anomaly mediation contribution since $F_\Phi = 0$ at the minimum, as shown in Eq.~\eqref{eq:fphi}.
In this model, it is possible to take \( M_{\tilde{b}} \sim M_{\tilde{w}} \) at the SUSY particle mass scale (the stop mass scale), which is necessary for coannihilation to work.  

It should be noted that to avoid a stau LSP or tachyonic stau in the parameter regions of interest, \( |B_1| \gg |B_5| \) is required. This is because large $|B_1|$ is needed to increase the stau masses. Therefore, \( B_1 / B_5 < 0 \) is necessary to explain $M_{\tilde b} \sim M_{\tilde w}$ ($g_1^2|B_1+B_5| \sim g_2^2 |B_5|$).

In this section and Sec.~\ref{sec:higgsino_wino}, $H_u$ and $H_d$ are taken to be bulk fields, in contrast to the brane-localized treatment adopted in Sec.~\ref{sec:bino_entropy} (see the discussion around Eq.~\eqref{eq:muterm4}). The Higgs $\mu$-term and $B$-term can be generated by including the following interaction terms,
\begin{eqnarray}
	\int d^4\theta \left( r_{ud} \tilde{h}(\rho) H_u H_d + h.c.+ \tilde{h}_{u}(\rho) |H_{u}|^2 +  \tilde{h}_{d}(\rho) |H_{d}|^2
	\right)
\end{eqnarray}
in $\mathcal{L}_0$. Here, we consider that $H_u$ and $H_d$ have profiles that are tilted toward $y=L$ due to the bulk mass terms. The effects are expressed by using $r_{ud}$. For simplicity we take $\tilde{h}_{u}=\tilde{h}_{d}=0$; nonzero $\tilde{h}_{u,d}$ can shift $m_A$, $\tan\beta$, $\mu$, and the Yukawa $A$-terms, but as long as $\tilde{h}_{u,d}$ are not too large, the impact on the neutralino dark matter phenomenology is small except in extreme regions of parameter space. Tuning $\mu$ down to $\mathcal{O}(1\,\mathrm{TeV})$ through this effect would in addition require substantial fine-tuning, which we do not pursue here.

The $\mu$-term and the Higgs B-term are given by
\begin{eqnarray}
	\mu = r_{ud} \tilde{h}' F_Z, \ \ 
	B_\mu = \frac{1}{2}r_{ud} \tilde{h}'' |F_Z|^2
\end{eqnarray}
Neglecting $\tilde{h}_{u}$ and $\tilde{h}_{d}$, we need a fine-tuning for $\tilde{h}''$. However, it is possible to avoid the significant fine-tuning with large $m_{H_d}^2$~\cite{Asano:2016oik,Nagai:2017tuf}.~\footnote{The fine-tuning to obtain the correct electroweak symmetry breaking scale can not be avoided.}

\paragraph{Bino-wino coannihilation}

\begin{figure}[htp]
	\centering
	%\hspace*{-5mm}
	\includegraphics[scale=0.8]{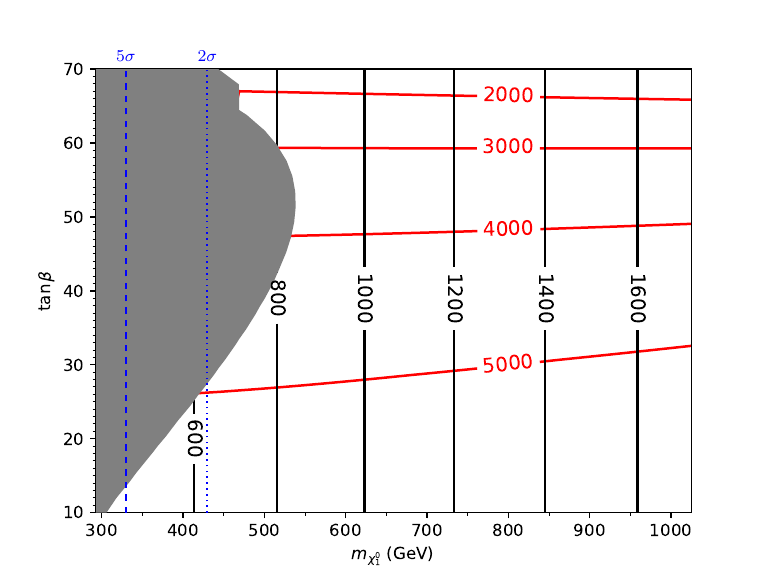}
	\caption{Bino-wino coannihilation scenario: the $m_{\chi_1^0}$\,-\,$\tan\beta$ plane for $B_3=1300$\, TeV. The black and red lines are contours of $m_{{\tilde \nu}_e} (\simeq m_{{\tilde e}_L} \simeq m_{{\tilde \mu}_L})$ and $m_A$ in units of GeV. The gray region is not consistent for our purpose since the stau is the LSP or tachyonic. The blue dashed and dotted vertical lines show the HL-LHC ($3\,{\rm ab}^{-1}$) $5\sigma$ discovery ($m_{\chi_1^0}=330$~GeV) and $2\sigma$ exclusion ($m_{\chi_1^0}=430$~GeV) reach for this scenario from Ref.~\cite{Duan:2018rls}.}
	\label{fig:1}
\end{figure}

In the case where the bino-like LSP with wino-like NLSP and the mass splitting is around 20-30 GeV, it is possible to explain the observed dark matter abundance~\cite{Harigaya:2014dwa}.
The bino, which tends to be overproduced, is efficiently annihilated through the annihilation of the wino-like NLSP.

For numerical calculation, we define the messenger scale as $M_{\rm mess}=\lambda_1 \vevx{X_1}=\lambda_3 \vevx{X_3}=\lambda_5 \vevx{X_5}$.
In fig.~\ref{fig:1}, we present the consistent parameter region for bino dark matter coannihilating with the wino. The gray-shaded region is excluded, as the stau becomes either the LSP or tachyonic. Additionally, we show the mass contours for the sneutrino (left-handed slepton) and the heavy Higgs bosons ($H$, $A$, $H^\pm$). The region for $\tan\beta > 70$ and $m_A<2$ TeV is very likely to be excluded~\cite{ATLAS:2020zms, Choi:2020wdq}. 
In the plot, we set $M_{\rm mess} = 3 \cdot 10^6~\text{GeV}$, $M_1 = -1.05 M_2~(M_2 > 0)$ at the messenger scale, and $B_3 = 1.3 \cdot 10^6~\text{GeV}$. The gluino and stop masses are approximately 9~TeV and 11~TeV, respectively, which are consistent with the observed Higgs boson mass.
Due to the coannihilation requirement, the mass splitting between the lightest neutralino ($\chi_1^0$) and the chargino ($\chi_1^\pm$) is small, with $m_{\chi_1^\pm} - m_{\chi_1^0} \approx 20\,\text{--}\,30~\text{GeV}$.

For SUSY searches in our model, the production modes of (a) $\chi_2^0 \chi_1^\pm$, (b) $\chi_1^+ \chi_1^-$, and (c) slepton pairs are particularly relevant. These processes lead to the following characteristic final states:  
(a) off-shell $W$ and $Z$ bosons,  
(b) off-shell $W^\pm$ bosons,  
(c) dilepton plus missing transverse energy.  
Based on a CMS search using 137 fb$^{-1}$ of Run 2 data~\cite{CMS:2021edw}, and current SUSY search results~\cite{ATLAS:2024lda, ATLAS:2024vnc}, $m_{\chi^0_2} \simeq m_{\chi^\pm_1}$ up to about 250 GeV for $\Delta M \sim 30$~GeV are already excluded. It means that almost all parameter region of this scenario remains consistent with experimental constraints at present.

To assess the testability of processes (a) and (b) at the HL-LHC, we simulate signal events for benchmark points with $m_{\chi_1^0} = 400$--$1000$ GeV, taking the mass splitting $\Delta M \equiv m_{\chi_1^\pm}-m_{\chi_1^0}$ that reproduces the observed relic abundance for each $m_{\chi_1^0}$, following the non-perturbative coannihilation result of Ref.~\cite{Harigaya:2014dwa}. Signal and the dominant backgrounds ($WZ$ for (a), $WW$ for (b), and $t\bar t$ for both) are generated at $\sqrt{s}=14$ TeV with {\tt MadGraph5\_aMC@NLO}~\cite{Alwall:2014hca}, showered with {\tt Pythia~8}~\cite{Sjostrand:2014zea}, and passed through a fast detector simulation with {\tt Delphes~3}~\cite{deFavereau:2013fsa} using an HL-LHC detector card. Both processes give leptons within a soft window set by the compressed spectrum, $5\,{\rm GeV}<p_T<\min(60\,{\rm GeV},\,1.5\,\Delta M)$, whose upper edge tracks the kinematic endpoint set by $\Delta M$; together with $E_T^{\rm miss}\gtrsim100$ GeV, this window is effective at separating the signal from $WW$/$WZ$/$t\bar t$, whose leptons and missing energy are typically harder. We find that the process (a), $\chi_2^0\chi_1^\pm\to3\ell+E_T^{\rm miss}$, is the more powerful of the two channels. The process (b), $\chi_1^+\chi_1^-\to\ell^+\ell^-+E_T^{\rm miss}$, provides a weaker cross-check.

\begin{figure}[t]
	\centering
	\includegraphics[width=0.48\textwidth]{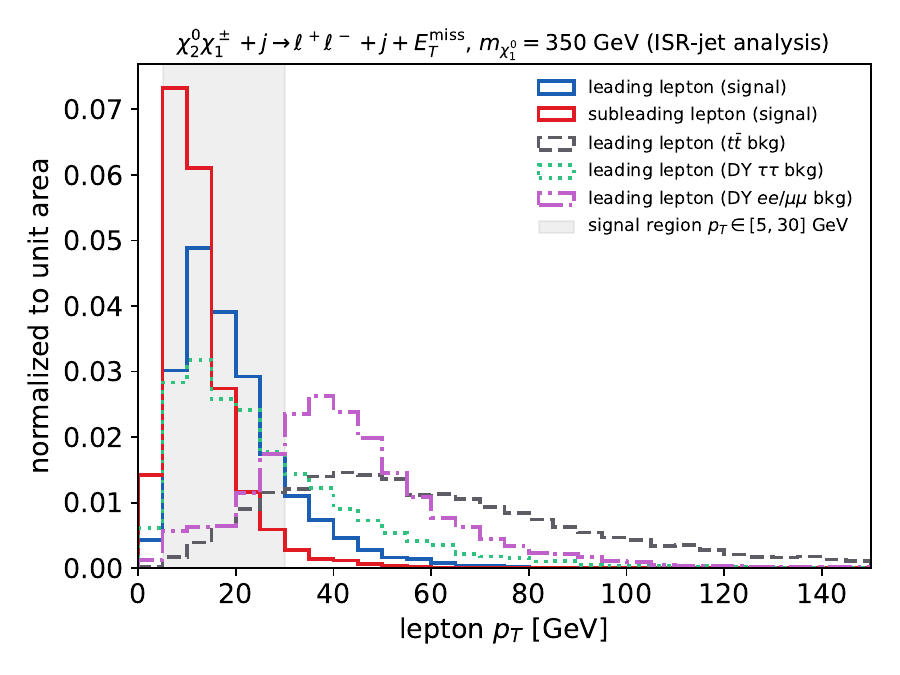}
	\includegraphics[width=0.48\textwidth]{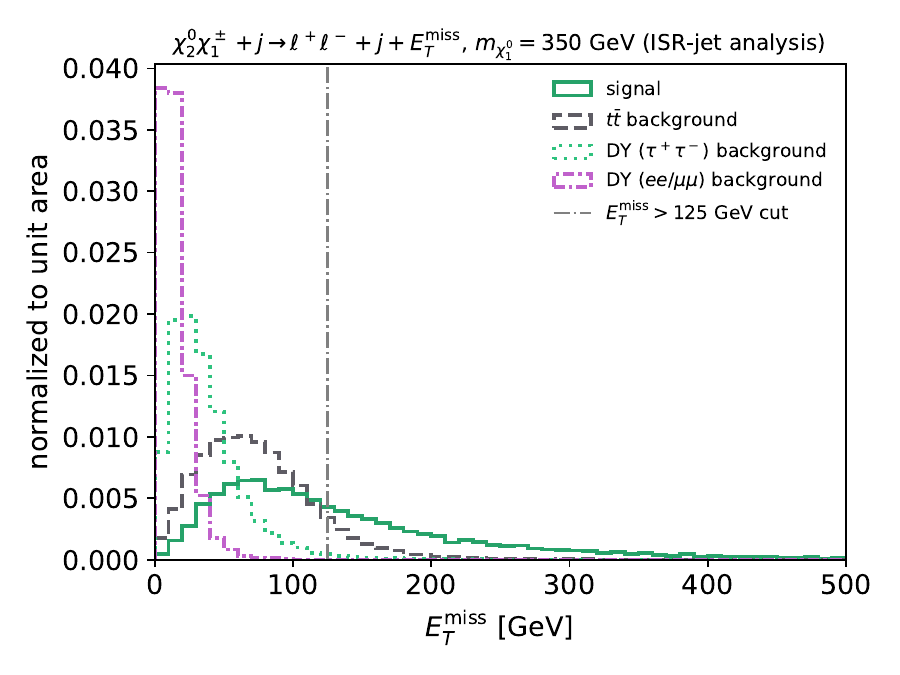}
	\caption{Cross-check of the jets-plus-soft-dilepton strategy of Ref.~\cite{Duan:2018rls} at $m_{\chi_1^0}=350$ GeV, $\Delta M=25$ GeV: lepton $p_T$ (left) and $E_T^{\rm miss}$ (right) distributions for the $\chi_2^0\chi_1^\pm+j$ signal and the $t\bar t$ and Drell-Yan backgrounds, before selection, normalized to unit area. We generate the two Drell-Yan channels separately: $\gamma^*/Z\to e^+e^-,\mu^+\mu^-$ (dash-dotted) and $\gamma^*/Z\to\tau^+\tau^-\to\ell^+\ell^-+4\nu$ (dotted). The signal leptons populate the soft window used in Ref.~\cite{Duan:2018rls} (shaded band, left panel) while the background leptons are harder; in $E_T^{\rm miss}$ (right panel), the direct $e^+e^-/\mu^+\mu^-$ channel peaks near zero as expected for a process with no genuine missing energy, while the $\tau^+\tau^-$ channel has a genuine high-$E_T^{\rm miss}$ tail from the four neutrinos in the $\tau$ decays, making it a non-negligible background even after the full selection of Ref.~\cite{Duan:2018rls}.}
	\label{fig:isr350}
\end{figure}

A naive leading-order, parton-shower-level significance estimate from this simulation omits trigger acceptance, additional backgrounds such as $t\bar{t}Z$ and nonprompt/misidentified leptons, and the finer kinematic selections (e.g.\ binning in the same-flavor dilepton invariant mass) used in a full experimental analysis, and should not be taken at face value. For more realistic estimates, we reproduced the jets-plus-soft-dilepton strategy of Ref.~\cite{Duan:2018rls} at $m_{\chi_1^0}=350$ GeV as a cross-check. To adequately populate the $E_T^{\rm miss}>125$ GeV tail with sufficient Monte Carlo statistics, Ref.~\cite{Duan:2018rls} performs the event generation with MLM-matching for the signal, Drell-Yan, and diboson samples up to one extra parton. As a simpler approximation, we generate the $\chi_2^0\chi_1^\pm+j$ signal and the Drell-Yan+jet background with a single hard parton at the matrix-element level ($p_T^j>40$ GeV) instead, while generating $t\bar t$ and $WW$ (a subdominant background in our cutflow) inclusively, and applied their full selection cuts:
\begin{itemize}
\item[] $E_T^{\rm miss}>125$ GeV, $n_j(p_{T,j}\ge 25~{\rm GeV}) \geq 1$ with a $b$-jet veto,
\item[] a soft opposite-sign same-flavor (OSSF) dilepton with $p_T\in[5,30]$ GeV,
\item[] $m_{\ell\ell}\in[4,50]$ GeV excluding the $\Upsilon$ window $m_{\ell\ell}\notin[9,10.5]$,
\item[] $H_T>100$ GeV, $E_T^{\rm miss}/H_T\in[0.6,1.4]$, and $M_T(\ell,E_T^{\rm miss})<70$ GeV.
\end{itemize}
Fig.~\ref{fig:isr350} shows the resulting lepton $p_T$ and $E_T^{\rm miss}$ distributions. Our cross-check gives $S/\sqrt{B}\approx7$ at $3\,{\rm ab}^{-1}$, consistent with a significance between the $5\sigma$ and $2\sigma$ boundaries quoted below for a point just above their $5\sigma$ reach. We regard this as confirming that our simulation chain and Ref.~\cite{Duan:2018rls}'s strategy are mutually consistent, but our own estimate still omits trigger acceptance, pile-up, and other backgrounds and selections used in a full experimental analysis, so we do not promote it to a quoted significance. We instead take the realistic HL-LHC reach from Ref.~\cite{Duan:2018rls}, which uses the same topology with full background estimation and a proper Poisson significance: at the HL-LHC ($3\,{\rm ab}^{-1}$) they find $m_{\chi_2^0}$ reaches up to $330$ GeV at $5\sigma$ and $430$ GeV at $2\sigma$, compared to $230$ GeV ($5\sigma$) and $310$ GeV ($2\sigma$) at $300\,{\rm fb}^{-1}$; independently, a CMS search using $137\,{\rm fb}^{-1}$ of Run~2 data already excludes $m_{\chi_2^0}\simeq m_{\chi_1^\pm}$ up to about $250$ GeV for $\Delta M\sim30$ GeV~\cite{CMS:2021edw}, consistent with this projection. We take the HL-LHC $5\sigma$--$2\sigma$ sensitivity range, $m_{\chi_1^0}\sim330$--$430$ GeV, as the realistic HL-LHC sensitivity of this scenario.

We have also examined channel (c), slepton pair production, and confirm quantitatively that it is subdominant to channel (b), using the same $m_{\chi_1^0}=350$ GeV benchmark point as the cross-check above. The left-handed slepton ($m_{\tilde e_L}=500$ GeV) decays directly to the LSP only $11.5\%$ of the time; $58.9\%$ of the decays proceed via the chargino and $29.5\%$ via the second neutralino. 
Since the chargino leg is itself compressed, as in channel (b), it does not produce a hard lepton at the primary vertex, so only the direct decay and the $\chi_2^0$-cascade feed a hard, prompt lepton per slepton leg; unlike channels (a) and (b), the resulting hard leptons face no soft-lepton trigger penalty. We generate the three corresponding exclusive final states (both legs direct, both legs via $\chi_2^0$, and the mixed combination) with {\tt MadGraph5\_aMC@NLO} and pass them through the same {\tt Pythia~8}+{\tt Delphes~3} chain as channels (a) and (b). Requiring two hard ($p_T>25$ GeV) opposite-sign same-flavor leptons and $E_T^{\rm miss}>200$ GeV, the combined selection efficiency is $15$--$23\%$ (lower than at $m_{\chi_1^0}=500$ GeV, where it reached $40$--$46\%$, because the smaller mass splittings at this point soften the leptons and the missing energy); combining the three channels (including the $\tilde e_L$ and $\tilde\mu_L$ contributions) gives $S\approx 13$ signal events at $3\,{\rm ab}^{-1}$, to be compared with $WW$ and $t\bar{t}$ backgrounds of $B\approx8.6\times10^3$ and $B\approx1.6\times10^5$ events, respectively, at the same selection ($t\bar t$ falls to $B\approx1.6\times10^4$ with a $b$-jet veto as in channel (b)); this gives $S/\sqrt{B}\approx0.03$--$0.08$, confirming that the slepton-pair channel contributes negligibly to the sensitivity discussed above.
\smallskip

We compute the spin-independent (SI) bino-proton cross section for the same seven benchmark points used in the collider study above, using {\tt micrOMEGAs 6.1.15}~\cite{Belanger:2001fz, Belanger:2004yn}. The LSP composition is confirmed to be $>99.99\%$ bino at every point, consistent with the tree-level diagonalization described above. Table~\ref{tb:directdetection} shows the resulting $\sigma_{\rm SI}$ together with its ratio to the current upper-bound on the SI cross section from the LUX-ZEPLIN (LZ) experiment~\cite{LZ:2024zvo},
\begin{eqnarray}
	\sigma_{\rm SI}^{\rm LZ} \lesssim 3 \times 10^{-11}\,{\rm pb}
	\ \left(\frac{m_{\chi_1^0}}{1\, {\rm TeV}} \right). \label{eq:lux_lz}
\end{eqnarray}
\begin{table}[t]
	\caption{Spin-independent bino-proton cross section $\sigma_{\rm SI}$, computed with {\tt micrOMEGAs 6.1.15} for the seven bino-wino coannihilation benchmark points of Sec.~\ref{sec:model}, and its ratio to the LZ upper bound of Eq.~\eqref{eq:lux_lz}.}
	\begin{center}
		\begin{tabular}{c c c}
			$m_{\chi_1^0}$ (GeV) & $\sigma_{\rm SI}$ (pb) & $\sigma_{\rm SI}/\sigma_{\rm SI}^{\rm LZ}$ \\
			\hline
			400  & $5.5\times10^{-15}$ & $4.6\times10^{-4}$ \\
			500  & $9.4\times10^{-16}$ & $6.3\times10^{-5}$ \\
			600  & $7.2\times10^{-17}$ & $4.0\times10^{-6}$ \\
			700  & $2.2\times10^{-15}$ & $1.1\times10^{-4}$ \\
			800  & $7.4\times10^{-15}$ & $3.1\times10^{-4}$ \\
			900  & $1.6\times10^{-14}$ & $5.8\times10^{-4}$ \\
			1000 & $2.7\times10^{-14}$ & $9.0\times10^{-4}$ \\
		\end{tabular}
	\end{center}
	\label{tb:directdetection}
\end{table}
This model therefore comfortably avoids the current LZ constraint, by three to six orders of magnitude across this mass range, and the predicted cross sections are also well below typical projections for the neutrino fog~\cite{OHare:2021utq}, making direct detection of this scenario very challenging even for next-generation experiments. The spin-dependent cross section is essentially constant over this mass range, $\sigma_{\rm SD}\approx(1.2\mathchar`-1.3)\times10^{-10}$ pb for both proton and neutron, and lies far below current bounds from PICO-60~\cite{PICO:2019vsc} and LZ~\cite{LZ:2024zvo}.

\section{Higgsino-like or wino-like dark matter} \label{sec:higgsino_wino}
In some regions of $B_1$, $B_3$, and $B_5$, the higgsino-like state or wino-like state becomes the LSP and a viable candidate for dark matter.  
Higgsino dark matter with a mass of approximately $1.1 \, \mathrm{TeV}$ and wino dark matter with a mass of approximately $2.8 \, \mathrm{TeV}$ can account for the observed dark matter density purely through thermal production~\cite{Hisano:2006nn, Hryczuk:2010zi, Hryczuk:2011vi, Bottaro:2021snn, Bottaro:2022one}.

Although these types of dark matter do not have significant mixing between gauginos and higgsinos, Higgs-mediated diagrams remain important for direct detection. Table \ref{tb:masses} presents the SUSY mass spectra for the higgsino-like LSP (P1) and the wino-like LSP (P2), along with the spin-independent cross section $\sigma_{\rm SI}$.  
In both cases, there are regions of parameter space where the LZ constraint in \eqref{eq:lux_lz} is satisfied. However, the cross section is generally larger compared to the earlier bino LSP scenario, meaning these regions are likely to be probed by future direct detection experiments with relative ease.

For wino-like dark matter, there might be tensions with indirect detection, depending on the dark matter density profile and the cosmic propagation model~\cite{Boveia:2022adi}.

\begin{table*}[t]
	\caption{Higgsino-like (P1) and wino-like (P2) dark matter scenario: the mass spectra and $\sigma_{\rm SI}$ (in pb). All masses are shown in units of TeV.}
	\begin{center}
		\begin{tabular}{c c c}
			& P1 & P2  \\
			\hline
			$M_{\rm mess}$          & $10^4$ & $3\times10^3$ \\
$B_5$                 & 2335 & 1050 \\
$B_1$                 & 8335 & 2000 \\
$B_3$                 & 0 & 2000 \\

			$\tan\beta$               & 8 & 7 \\
			\hline
			$\tilde{g}$                & 14.4 & 18.2 \\
			$\tilde{t}$                & 20.0 & 19.0  \\
			${\tilde{\tau}_1}$  & 11.2 & 4.06  \\
			$({\chi^0_1},{\chi^\pm_1})$
			& (1.080, 1.081) & (2.827, 2.827)  \\
			$\mu$      & 1.046 & 8.273 \\
			\hline
			$\sigma_{\rm SI} ({\rm pb})$ &
$2.7 \cdot 10^{-11}$ & $10^{-11}$
		\end{tabular}
	\end{center}
	\label{tb:masses}
\end{table*}%

\section{Bino-like dark matter with entropy production} \label{sec:bino_entropy}

As an alternative to the coannihilation scenario, in this section, we consider a mechanism where the over-abundant bino-like neutralino is diluted by additional entropy production, thereby explaining the observed dark matter abundance. For this purpose, we utilize the mechanism introduced in Refs.~\cite{Fujii:2002fv, Hamaguchi:2014sea}.

Instead of the superpotential presented in Eq.~\eqref{eq:mess1}, we consider:  
\begin{eqnarray}  
	W = \lambda_5 X_5 \Psi_5 \bar{\Psi}_5 + \frac{\kappa_5}{3}X_5^3 + W_{\rm mix},  
\end{eqnarray}  
where $W_{\rm mix}$ represents the mixing between the messenger field and matter fields.

Neglecting $W_{\rm mix}$, the lightest state of the messenger field, typically the scalar component of the weak doublet, becomes stable and dominates the energy density of the universe as it cools. However, by including a small mixing term in $W_{\rm mix}$, the messenger field decays into MSSM particles at a late time. Because of this late decay, the messenger temporarily dominates the energy density of the universe, producing entropy during its decay and diluting the bino abundance.

The small mixing term is given by  
\begin{eqnarray}  
	W_{\rm mix} = \tilde{m} H_{\bf 5} \Psi_{\bar{\bf 5}} \ni \tilde{m} H_u \Psi_{\rm weak} .
	\label{eq:wmix}
\end{eqnarray}
The smallness of $\tilde{m}$ can be explained by certain symmetries and their breaking (spurions): as noted in Sec.~\ref{sec:model}, the discrete symmetry $Z_{3,5}$ that forbids messenger-matter mixing in the scenarios of Secs.~\ref{sec:model}--\ref{sec:higgsino_wino} is broken here by a small amount to generate $\tilde{m}$. With the mixing term, the lightest scalar messenger predominantly decays into a pair of bottom quarks. The decay rate is given by  
\begin{eqnarray}  
	\Gamma_{\rm mess} \simeq \frac{1}{8\pi} y_{b}^2 \left( \frac{\tilde{m}}{M_{\rm mess}} \right)^2 M_{\rm mess},  
\end{eqnarray}  
where $M_{\rm mess} \equiv \lambda_5 \langle X_5 \rangle$.

The thermal abundance of the messenger (defined as $n_{\rm mess}/s$) after freeze-out is given by~\cite{Dimopoulos:1996gy, Han:1997wn}:  
\begin{eqnarray}  
	Y_{\rm mess} \approx 3.7 \times 10^{-8} \left( \frac{M_{\rm mess}}{10^8\, \text{GeV}} \right).  
\end{eqnarray}  
If the scalar messenger's decay is sufficiently late, it can dominate the energy density of the universe before decaying. In this case, it produces an entropy, thereby diluting the dark matter energy density. The resulting dilution factor is  
\begin{eqnarray}  
	\Delta = \frac{s_{\rm after}}{s_{\rm before}} = \frac{4}{3} \frac{M_{\rm mess} Y_{\rm mess}}{T_d},  
\end{eqnarray}  
where $T_d \sim \sqrt{\Gamma_{\rm mess} M_P}$ is the decay temperature. To ensure compatibility with BBN and dark matter freeze-out, we require $\mathcal{O}(1\, \text{MeV}) < T_d < T_{\rm fo} $, where $ T_{\rm fo} \sim m_{\chi_1^0}/20 $ represents the freeze-out temperature.  

With this dilution mechanism, the observed dark matter abundance is expressed as  
\begin{eqnarray}  
	\Omega_{\rm obs} = \frac{1}{\Delta} \Omega_{\chi_1^0}.  
\end{eqnarray}  
Typically, a dilution factor of $\Delta = \mathcal{O}(100)$ is required. To satisfy this condition, the decay temperature must lie in the range $ 10~\text{MeV} < T_d < 100~\text{GeV} $, which implies that the messenger scale $ M_{\rm mess} $ should be within $ \mathcal{O}(10^7)~\text{GeV} $ and $ \mathcal{O}(10^{10})~\text{GeV} $.

Unlike in Secs.~\ref{sec:model} and \ref{sec:higgsino_wino}, where $H_u$ and $H_d$ are bulk fields, here we assume for simplicity that $H_u$ and $H_d$ are localized on the brane at $y=L$, together with the MSSM matter fields. In this setup, there exists a $\mu$-term in the superpotential given by:
\begin{eqnarray}
	\mathcal{L}_L \ni \int d^2\theta \, \mu H_u H_d + h.c.
	\label{eq:muterm4}
\end{eqnarray}
At tree level, there is no $B$-term, but it is radiatively generated. Consequently, $ \tan\beta $ is not a free parameter but rather a prediction of the model.

\begin{figure}[htp]
	\centering
	%\hspace*{-5mm}
	\includegraphics[scale=0.8]{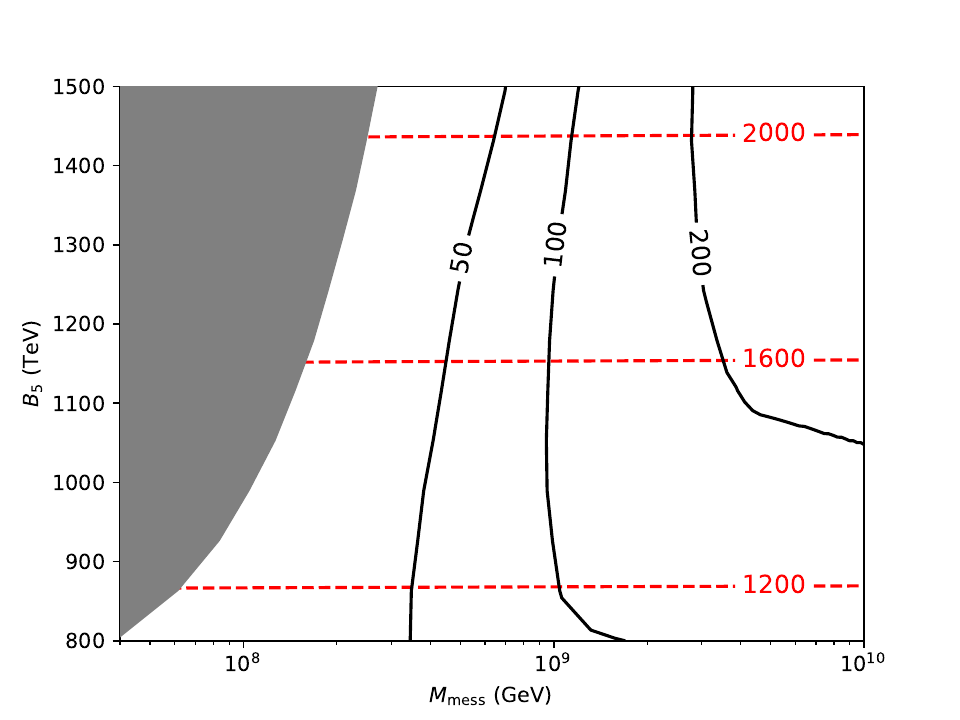}
	\caption{Entropy-diluted bino dark matter scenario: the required dilution factor $\Delta$ (black solid) and $m_{\chi_1^0}$ in units of GeV (red dashed). In the gray region, the stau is the LSP or tachyonic.}
	\label{fig:2}
\end{figure}

\begin{figure}[htp]
	\centering
	%\hspace*{-5mm}
	\includegraphics[scale=0.8]{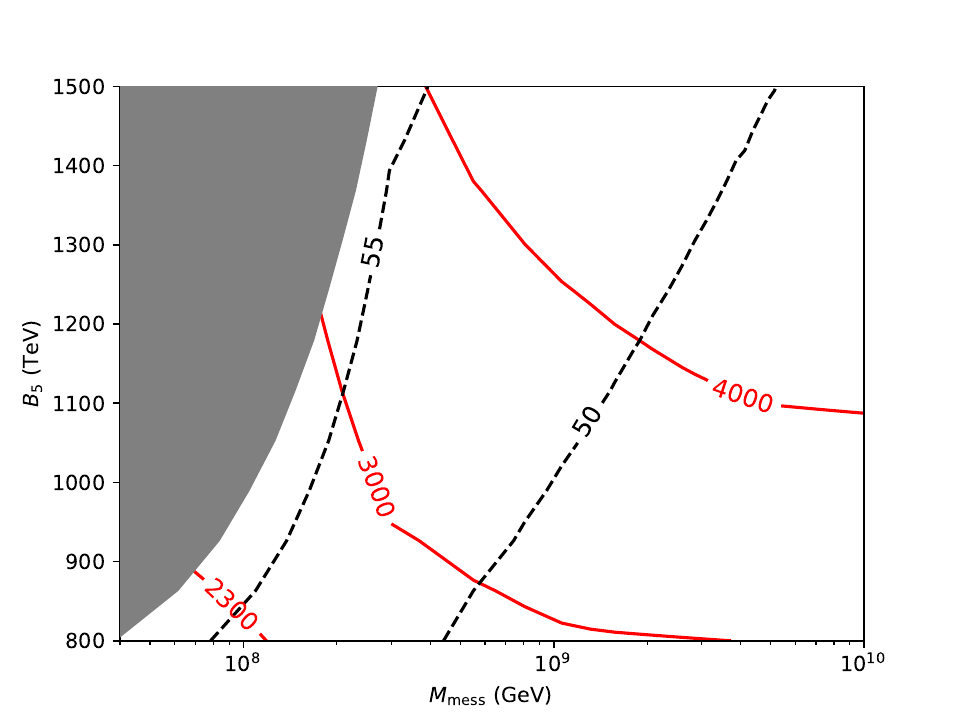}
	\caption{Entropy-diluted bino dark matter scenario: the predicted $\tan\beta$ (black dashed) and $m_{A}$ in units of GeV (red solid).}
	\label{fig:3}
\end{figure}

In fig.~\ref{fig:2}, we plot the contours of the required dilution factors, while fig.~\ref{fig:3} illustrates the predicted $\tan\beta$ and the masses of the heavy Higgs bosons ($H$, $A$, $H^\pm$). As discussed in \cite{Choi:2020wdq}, there is an opportunity to test the heavy Higgs bosons through searches for heavy Higgs decays into the final state $\tau^+ \tau^-$.  The gluino mass approximately ranges from 5 to 9 TeV, while the stop mass ranges from around 6 to 11 TeV.
We compute the SI cross section between the bino and proton with {\tt micrOMEGAs 6.1.15} for fixed $B_5=1300$ TeV, varying the messenger scale $M_{\rm mess}$ over the range relevant for the dilution mechanism (the corresponding $\tan\beta$, shown in Table~\ref{tb:dilutedSI}, is fixed at each point by the tree-level condition $B_\mu=0$, consistent with the prediction of the model and with fig.~\ref{fig:3}). For $M_{\rm mess}\lesssim3\times10^8$ GeV at this $B_5$, the stau becomes lighter than the bino and this region is excluded, as reflected in the gray region of fig.~\ref{fig:2}; Table~\ref{tb:dilutedSI} is restricted to the allowed region. As shown in the table, $\sigma_{\rm SI}$ decreases mildly with $M_{\rm mess}$: since $H_u$ and $H_d$ are brane-localized in this scenario, $\mu$ is fixed by radiative electroweak symmetry breaking rather than by an independent parameter, and $|\mu|$ grows slowly with $M_{\rm mess}$, suppressing the bino-higgsino mixing responsible for the SI coupling. Across this range, $\sigma_{\rm SI}$ remains a few percent of the current LZ bound of Eq.~\eqref{eq:lux_lz}, considerably closer to the current experimental sensitivity than the bino-wino coannihilation scenario of Sec.~\ref{sec:model} (Table~\ref{tb:directdetection}), and this scenario may therefore be probed by near-future direct detection experiments.
\begin{table}[t]
	\caption{Entropy-diluted bino dark matter scenario: dependence of the spin-independent bino-proton cross section on the messenger scale $M_{\rm mess}$, for fixed $B_5=1300$ TeV, with $\tan\beta$ at each point fixed by the tree-level condition $B_\mu=0$, computed with {\tt micrOMEGAs 6.1.15}.}
	\begin{center}
		\begin{tabular}{c c c c c c}
			$M_{\rm mess}$ (GeV) & $\tan\beta$ & $m_{\chi_1^0}$ (GeV) & $\mu$ (GeV) & $\sigma_{\rm SI}$ (pb) & $\sigma_{\rm SI}/\sigma_{\rm SI}^{\rm LZ}$ \\
			\hline
			$5.5\times10^8$  & 53.1 & 1815 & $4692$ & $1.9\times10^{-12}$ & $3.6\times10^{-2}$ \\
			$1.1\times10^9$  & 51.8 & 1814 & $4769$ & $1.8\times10^{-12}$ & $3.3\times10^{-2}$ \\
			$1.6\times10^9$  & 51.1 & 1813 & $4810$ & $1.7\times10^{-12}$ & $3.1\times10^{-2}$ \\
			$2.3\times10^9$  & 50.4 & 1813 & $4849$ & $1.6\times10^{-12}$ & $3.0\times10^{-2}$ \\
			$3.4\times10^9$  & 49.8 & 1813 & $4882$ & $1.6\times10^{-12}$ & $2.9\times10^{-2}$ \\
			$5.1\times10^9$  & 49.0 & 1812 & $4918$ & $1.5\times10^{-12}$ & $2.8\times10^{-2}$ \\
			$10^{10}$        & 48.1 & 1812 & $4967$ & $1.5\times10^{-12}$ & $2.7\times10^{-2}$ \\
		\end{tabular}
	\end{center}
	\label{tb:dilutedSI}
\end{table}

Finally, the scalar component of $\rho$ has a mass of the order of $m_{3/2} M_P^3/M_*^3$ and decays into a pair of gravitinos. If the initial abundance of $\rho$ is too large, the LSP abundance generated from gravitino decay may exceed acceptable limits unless the reheating temperature is kept very low. To address this issue, the inflaton field can be placed on the $ y=0 $ brane, suppressing the initial amplitude of $\rho$ through the enhanced coupling between $ Z $ and the inflaton~\cite{Linde:1996cx, Nakayama:2011zy, Nakayama:2012mf}.~\footnote{The K\"ahler potential includes a term of the form $ \sim \frac{M_P^2}{M_*^2} \frac{|Z|^2 |I|^2}{M_*^2} $ after the canonical normalization of $ Z $, where $ I $ denotes the inflaton field.} This setup allows for a reheating temperature higher than $ M_{\rm mess} $. For the scalar component of the imaginary part of $Z$, its mass and behavior depend on the breaking of the shift symmetry. We leave a detailed discussion of this point for future work.

\section{Radion stabilization} \label{sec:radion}

We consider a supersymmetric version of Goldberger--Wise radion
stabilization governed by the action%
\cite{Goldberger:1999uk,Maru:2003mq,Eto:2004yk}
\begin{align}
	S &= S_{\text{bulk}}[H,H^c,T] + S_0[H,Z] + S_\pi[H,Q,T] \,.
\end{align}
Here, $T$ denotes the radion chiral multiplet and $(H,H^c)$ is a bulk hypermultiplet
expressed in terms of 4D $\mathcal{N}=1$ chiral superfields~\cite{Marti:2001iw,PaccettiCorreia:2004suu},
with $H$ assigned $Z_2$-even parity and $H^c$ assigned $Z_2$-odd parity,
so that only $H$ has a zero mode. Throughout this section we use the
dimensionless orbifold coordinate $\hat y\in[0,\pi]$, reserving the symbol $y$
for the dimensionful coordinate $y\in[0,L]$ of
Secs.~\ref{sec:model}--\ref{sec:bino_entropy}; the two are related by
$y=R\hat y$, where $R$ is the compactification radius fixed by the radion
VEV (see below). The dimensionless parameter $c$, appearing
below in units of the AdS curvature $k$, is the bulk mass of the
hypermultiplet; it sets the zero mode's localization profile along $\hat y$,
which is UV-peaked ($\hat y=0$) for $c>1/2$ and IR-peaked ($\hat y=\pi$) for
$c<1/2$, and reduces to the bulk mass parameter of Ref.~\cite{Maru:2003mq}
with the same sign convention.
The fields $Z$ and $Q$ are localized on the UV brane at $\hat y=0$ and
the IR brane at $\hat y=\pi$ (identified below with the $y=0$ and $y=L$ branes of
Sec.~\ref{sec:model}, respectively); $Z$ is the same SUSY-breaking field
introduced in Sec.~\ref{sec:model}, with $\rho\equiv Z+Z^\dagger$ and $g(\rho)$ its Kähler
function as in Eq.~\eqref{eq:gravsusy}. For brevity we write
$\Omega_Z(\rho)\equiv M_*^2\,g(\rho)$ for the combination that appears below. The field
$Q$ is a representative of the MSSM and messenger fields, which are localized
on the IR brane; it is kept explicit to show that these fields acquire no soft
SUSY-breaking parameters.
We work in Planck units ($M_P=1$) and take the AdS curvature
$k\sim0.1\,M_P$~\cite{Randall:1999ee}; the size of the warp factor
required by sequestering is discussed at the end of this section.
The bulk and brane actions, written in terms of the conformal compensator
$\Phi$, are given by
%
% --- Bulk section ---
%
\begin{align}
	S_{\text{bulk}} &= \int d^4x \int_0^{\pi} d\hat y\;
	\Bigg[
	\int d^4\theta\; \Phi^\dagger \Phi\,
	\frac{T+T^\dagger}{2}\,
	e^{-k|\hat y|\,(T+T^\dagger)}
	\left( H^\dagger H + H^{c} H^{c\dagger} \right) \nonumber
	\\
	&\qquad
	+ \left\{
	\int d^2\theta\; \Phi^3 e^{- 3 k|\hat y|\,T}\, H^c
	\left( \partial_{\hat y} - \left(\frac{3}{2}-c\right)k\,T\,\mathrm{sgn}(\hat y) \right) H
	+ \text{h.c.}
	\right\}
	\Bigg] .
	\label{eq:bulk_action}
\end{align}
Integrating by parts in $\hat y$ brings this operator to the form
$H\big(-\partial_{\hat y}+(\tfrac32+c)kT\,\mathrm{sgn}(\hat y)\big)H^c$, which
is the bulk hypermultiplet action of Ref.~\cite{Maru:2003mq}.
The corresponding brane actions are
% --- Z on the \hat y = 0 brane ---
%
\begin{align}
	S_0 &= \int d^4x \int_0^{\pi} d\hat y\;
	\delta(\hat y)\Bigg[
	\int d^4\theta\; \Phi^\dagger \Phi\, \Omega_Z(\rho)
	\nonumber	\\
	&\qquad\qquad
	+ \left\{
	\int d^2\theta\;  \Phi^3  W_0(H)
	+ \text{h.c.}
	\right\}
	\Bigg] ,
\end{align}
\begin{align}
	W_0(H) = J_0\,H + \mathcal{C},
\end{align}
where $\mathcal{C}$ is a constant term.
We assume the action possesses a shift symmetry $Z \to Z + i \alpha$,
allowing $Z$ to act as a SUSY-breaking field with a shift-symmetric
K\"ahler potential and a vanishing superpotential,%
%
% --- Q on the \hat y = \pi brane ---
%
\begin{align}
	S_\pi &= \int d^4x \int_0^{\pi} d\hat y\;
	\delta(\hat y-\pi)\Bigg[
	\int d^4\theta\; \Phi^\dagger \Phi e^{-k\pi (T+T^\dagger)} Q^\dagger Q
	\nonumber	\\
	&\qquad\qquad
	+ \left\{
	\int d^2\theta\;  \Phi^3  W_\pi(H,Q)\, e^{-3k\pi T}
	+ \text{h.c.}
	\right\}
	\Bigg] ,
\end{align}
\begin{align}
	W_\pi(H,Q) = J_\pi\,H + W(Q).
\end{align}
The 5D Planck mass $M_5$ entering the frame function below is related to
$M_P$ through the standard Kaluza--Klein reduction of the 5D
Einstein--Hilbert action, $M_P^2=\dfrac{M_5^3}{k}\left(1-e^{-2k\pi R}\right)$.
Here $R$ is the compactification radius fixed by the radion VEV,
$\langle T\rangle=R$, so that $\tau_0\equiv\langle T+T^\dagger\rangle=2R$ and
the warp factors of the two descriptions agree, $e^{-k\pi\tau_0}=e^{-2k\pi R}$;
equivalently, the frame function introduced below obeys
$\Omega_{\rm grav}(\tau_0)=-3M_P^2$.
Since the metric component along the extra dimension is unwarped,
$g_{\hat y\hat y}=R^2$ for the dimensionless coordinate $\hat y\in[0,\pi]$ used
here, the proper distance between the two branes is simply $\pi R$; identifying
this with the brane separation $L$ of Sec.~\ref{sec:model} gives $L=\pi R$, so
that $\hat y\in[0,\pi]$ and the UV/IR branes at its endpoints correspond to the
dimensionful coordinate $y=R\hat y\in[0,L]$ and the $y=0$/$y=L$ branes of
Secs.~\ref{sec:model}--\ref{sec:bino_entropy}.

By integrating out the bulk hypermultiplet using the boundary conditions
induced by $J_0$ and $J_\pi$, and focusing on the light modes, following the
same steps as Ref.~\cite{Maru:2003mq},
we obtain an effective 4D $\mathcal{N}=1$ supergravity description in terms of
the radion $T$, the zero mode $h$ of $H$ (the odd field $H^c$ has no zero
mode), the SUSY-breaking field $Z$, and the IR-localized
field $Q$~
\cite{Marti:2001iw,PaccettiCorreia:2004suu,PaccettiCorreia:2006rpo}.
The 4D effective theory can be cast in the standard form
\begin{align}
	\mathcal{L}_{\rm 4D}
	&=
	\int d^4\theta\,
	\Phi^\dagger\Phi\,
	\Omega
	+
	\Big[
	\int d^2\theta\,
	\Phi^3\,W
	+ {\rm h.c.}
	\Big],
\end{align}
whose scalar potential, after eliminating the compensator, takes the
standard $\mathcal{N}=1$ supergravity form (in units $M_P=1$)
\begin{align}
	V = e^{K}\Big(K^{i\bar j}D_iW\,\overline{D_jW}-3|W|^2\Big),
	\qquad
	D_iW \equiv \partial_iW + (\partial_iK)\,W,
	\qquad
	F^i = -e^{K/2}K^{i\bar j}\overline{D_jW},
\end{align}
with the (frame) K\"ahler potential ($\Omega = -3 e^{-K/3}$)
\begin{align}
	\Omega(T,T^\dagger,h,h^\dagger,Z,Z^\dagger,Q,Q^\dagger)
	&=
	\Omega_{\rm grav}(T,T^\dagger)
	+ K_h(T,T^\dagger)\,h\,h^\dagger
	+ \Omega_Z(\rho)
	+ e^{-k\pi (T+T^\dagger)}\,Q^\dagger Q,
	\\[4pt]
	\Omega_{\rm grav}(T,T^\dagger) &= -\frac{3M_5^3}{k}\Big[1-e^{-(T+T^\dagger)k\pi}\Big],
	\\[4pt]
	K_h(T,T^\dagger) &= \frac{e^{(1/2-c)(T+T^\dagger)k\pi}-1}{(1-2c)\,k},
	\label{eq:radion_omega}
\end{align}
and the superpotential
\begin{align}
	W(T,h,Q)
	&= \mathcal{C} + h\,\Delta(T) + e^{-3 k\pi T}\,W(Q),
	\qquad
	\Delta(T) \equiv J_0 - J_\pi\, e^{-(c+3/2)k\pi T},
	\label{eq:radion_W}
\end{align}
where $\mathcal{C}$ is the same UV-brane constant as in Eq.~\eqref{eq:gravsusy};
it does not affect $D_hW|_{h=0}=\Delta(T)$, since it is $h$-independent,
but it does source $D_TW$ at $h=0$, as discussed below.
Here $\Omega_{\rm grav}(T,T^\dagger)$ is the frame function generated by reducing the bulk
gravity multiplet~\cite{Maru:2003mq} (denoted $f(T,T^\dagger)$ there; we
rename it here to avoid a clash with the unrelated function $f(\rho)$ of
Sec.~\ref{sec:model}), and $K_h(T,T^\dagger)$ is the induced
K\"ahler metric of the light mode $h$; both follow directly from the corrected
bulk action~\eqref{eq:bulk_action}, so that this 4D EFT reduces exactly to
that of Ref.~\cite{Maru:2003mq} when $Z,Q\to0$, with the same $c$. The K\"ahler
term $e^{-k\pi(T+T^\dagger)}Q^\dagger Q$ and the superpotential term
$e^{-3k\pi T}W(Q)$ involve the radion but not the SUSY-breaking field $\rho$, so
$F_Z$ generates no soft terms for $Q$; their radion dependence would do so only
through $F^T$ and $F_\Phi$, which both vanish at the minimum (shown below).
Since $Q$ has a vanishing VEV, we set $Q=0$ in the explicit derivation below.

Since $\Omega$ depends on $h$ only through $|h|^2$ and $W$ is linear in $h$,
$h=0$ extremizes the potential at every $T$, and the light-field content
relevant for radion stabilization reduces to $(T,h)$. At $h=0$ the Kähler
metric is block-diagonal,
\begin{align}
	K_{T\bar T}\big|_{h=0} = \frac{3(k\pi)^2}{4\sinh^2\!\big(\tfrac{k\pi(T+T^\dagger)}{2}\big)},
	\qquad
	K_{h\bar h}\big|_{h=0} = -\frac{3\,K_h(T,T^\dagger)}{\Omega_{\rm grav}(T,T^\dagger)}.
\end{align}
Keeping $\Omega_Z(\rho)$ explicit would instead give
$K_{h\bar h}|_{h=0}=-3K_h/(\Omega_{\rm grav}+\Omega_Z)$; we have dropped
$\Omega_Z$ relative to $\Omega_{\rm grav}$ in the denominator, which is the
same $M_*\ll M_P$ approximation already used in Sec.~\ref{sec:model} ($|\phi|^2\approx1$
for $M_*/M_P\sim0.1$), since $\Omega_Z/\Omega_{\rm grav}\sim(M_*/M_P)^2$ is
parametrically small. To this order the radion potential and mass derived
below coincide with the pure Maru--Okada result~\cite{Maru:2003mq}, up to
$\mathcal{O}\big((M_*/M_P)^2\big)$ corrections from $\Omega_Z$.
The radion does not carry a SUSY-breaking $F$-term, $F^T=0$ at the minimum,
even though $D_TW$ itself does not vanish. In Ref.~\cite{Maru:2003mq} the
superpotential carries no constant term, so that $\langle W\rangle=0$ at the
SUSY minimum and the gravitino mass has to be supplied by a separate hidden
sector, whose breaking reaches the radion only through the compensator;
$F^T$ then vanishes only to first order in $F_\Phi$, and only after the light
mode is displaced by $\delta h\propto F_\Phi$, leaving anomaly mediation as
the dominant transmission channel. Here, instead, the constant $\mathcal{C}$
that generates $m_{3/2}$ sits directly in the superpotential, and $F^T=0$ and
$F_\Phi=0$ both hold exactly at $h=0$ once $Z$ is included, as we now show. Using $W=\mathcal{C}+h\Delta(T)$ at $Q=0$ from Eq.~\eqref{eq:radion_W}
($\mathcal{C}\neq0$, since it also sources $Z$'s SUSY breaking,
$F_Z\propto1/\langle g'(\rho)\rangle$ of Sec.~\ref{sec:model}) and
$\Omega=\Omega_{\rm grav}(T,T^\dagger)+\Omega_Z(\rho)$ from
Eq.~\eqref{eq:radion_omega}, one finds
\begin{align}
	D_TW=-\frac{3\mathcal C\,\Omega_{\rm grav}'}{\Omega},\qquad
	D_ZW=-\frac{3\mathcal C\,\Omega_Z'}{\Omega},
\end{align}
both individually nonzero, where primes denote derivatives with respect to
$T+T^\dagger$ (for $\Omega_{\rm grav}$) or $\rho$ (for $\Omega_Z$). Although
$\Omega$ is additively separable in $T+T^\dagger$ and $\rho$, the Kähler
metric is not: it has a nonzero off-diagonal component
$K_{T\bar Z}=3\,\Omega_{\rm grav}'\Omega_Z'/\Omega^2$ mixing $T$ and $Z$, so
that $F^T=-e^{K/2}\big(K^{T\bar T}\overline{D_TW}+K^{T\bar Z}\overline{D_ZW}\big)$
with the full (non-diagonal) inverse metric. A direct computation gives the
exact identity
\begin{align}
	K_{Z\bar Z}\,\overline{D_TW}-K_{T\bar Z}\,\overline{D_ZW}
	= \frac{9\,\mathcal C\,\Omega_{\rm grav}'\,\Omega_Z''}{\Omega^2},
\end{align}
valid for any $\Omega_Z(\rho)$, so that $F^T\propto\Omega_Z''(\rho)$ exactly:
the individually nonzero $D_TW,D_ZW$ cancel completely in $F^T$ whenever
$\Omega_Z''(\rho)=0$. Explicitly,
\begin{align}
	F^T = \frac{3\sqrt{3}\,\mathcal C\,\Omega_{\rm grav}'\,\Omega_Z''}
	{\sqrt{-\Omega}\,\big[\Omega\,\Omega_{\rm grav}''\Omega_Z''-(\Omega_{\rm grav}')^2\Omega_Z''-\Omega_{\rm grav}''(\Omega_Z')^2\big]}.
\end{align}
Extending the scalar potential to include $Z$ likewise gives, at $h=0$,
\begin{align}
	V(T,Z) = \frac{81\,\mathcal C^2\,\Omega_{\rm grav}''\,\Omega_Z''}
	{\Omega^2\big[\Omega\,\Omega_{\rm grav}''\Omega_Z''-(\Omega_{\rm grav}')^2\Omega_Z''-\Omega_{\rm grav}''(\Omega_Z')^2\big]},
\end{align}
where primes on $\Omega_{\rm grav}$ denote derivatives with respect to
$T+T^\dagger$ and primes on $\Omega_Z$ derivatives with respect to $\rho$.
Requiring a vanishing cosmological constant, $V=0$ with $\mathcal C\neq0$,
gives $\Omega_{\rm grav}''\,\Omega_Z''=0$. Since
$\Omega_{\rm grav}''(T,T^\dagger)=3\pi^2kM_5^3e^{-(T+T^\dagger)k\pi}$ never
vanishes, this forces $\Omega_Z''(\rho)=M_*^2g''(\rho)=0$, i.e.\
$g''(\rho)=0$, at the minimum -- exactly the condition already obtained in
Sec.~\ref{sec:model} (Eq.~\eqref{eq:fphi}) from the $(Z,\Phi)$ system alone, and unchanged
by including $T$. Substituting $g''(\rho_0)=0$ gives $F^T=0$ identically
while $F_Z\propto1/g'(\rho_0)\neq0$ survives: the radion sector remains
exactly supersymmetric even though the full vacuum breaks SUSY.
Since $W=\mathcal{C}$ is constant, $V=-3\phi^3F_\Phi\mathcal{C}$ is the same
relation used in Sec.~\ref{sec:model} (Eq.~\eqref{eq:fphi}); hence $F_\Phi=0$
follows from $V(T,Z)=0$ above just as directly as $F^T=0$ does, at the same
$\Omega_Z''(\rho_0)=0$. Anomaly
mediation therefore remains absent once the radion is included, as in Sec.~\ref{sec:model}.
The radion potential itself is then governed by the $h$-direction alone,
with $Z$ fixed at its own vacuum, where
\begin{align}
	D_h W\big|_{h=0} = \Delta(T).
\end{align}
The scalar potential is then
\begin{align}
	V(T,T^\dagger) &= e^{K}K^{h\bar h}\big|D_hW\big|^2\Big|_{h=0} \notag\\
	&= \frac{k^3(2c-1)}{M_5^6}\,e^{(2c+3)k\pi(T+T^\dagger)/2}\,\Delta(T)\,\Delta(T^\dagger) \notag\\
	&\quad\times\Big[\big(e^{k\pi(T+T^\dagger)}-1\big)^2\big(e^{(2c-1)k\pi(T+T^\dagger)/2}-1\big)\Big]^{-1}.
\end{align}
Restricting to a real radion field, $\tau \equiv T+T^\dagger$, this becomes
\begin{align}
	V(\tau) &= \frac{k^3(2c-1)}{M_5^6}\,
	\frac{e^{k\pi\tau(2c+3)/2}}
	{\big(e^{k\pi\tau}-1\big)^2\big(e^{k\pi\tau(2c-1)/2}-1\big)}\,
	\Delta(\tau/2)^2.
	\label{eq:radion_potential}
\end{align}
The SUSY vacuum is at $\Delta(\tau_0/2)=0$, i.e.
\begin{align}
	J_0 = J_\pi\, e^{-k\pi\tau_0(2c+3)/4},
	\qquad h=0,
\end{align}
where $V(\tau_0)=0$. Since $V(\tau)\propto\Delta(\tau/2)^2$ and
$\Delta(\tau_0/2)=0$, the extremum condition $V'(\tau_0)=0$ is automatically
satisfied, and the second derivative reduces to
\begin{align}
	V''(\tau_0) &= \frac{k^3(k\pi)^2(2c-1)(2c+3)^2}{8\,M_5^6}\,
	\frac{J_\pi^2}{\big(e^{k\pi\tau_0}-1\big)^2\big(e^{k\pi\tau_0(2c-1)/2}-1\big)}.
\end{align}
Canonically normalizing the radion using $K_{T\bar T}(\tau_0)$, the radion
mass is
\begin{align}
	m^2_{\rm radion} = \frac{2\,V''(\tau_0)}{K_{T\bar T}(\tau_0)}
	= \frac{J_\pi^2\,k^3\,(2c-1)(2c+3)^2}{12\,M_5^6}\,
	\frac{1}{e^{k\pi\tau_0}\big(e^{k\pi\tau_0(2c-1)/2}-1\big)},
	\label{eq:radion_mass}
\end{align}
which is positive for all $c\neq1/2$: the factor $(2c-1)$ and
$\big(e^{k\pi\tau_0(2c-1)/2}-1\big)$ always share the same sign for
$\tau_0>0$.\footnote{Ref.~\cite{Maru:2003mq} normalizes the radion kinetic
term with the frame function $\partial_T\partial_{T^\dagger}\Omega_{\rm grav}$
rather than with the K\"ahler metric $K_{T\bar T}$; in units $M_P=1$ the
two are related by $K_{T\bar T}=\partial_T\partial_{T^\dagger}
\Omega_{\rm grav}/(1-e^{-k\pi\tau_0})$, so that Eq.~\eqref{eq:radion_mass} is
smaller than their estimate by $1-e^{-2k\pi R}$, a $\lesssim0.2\%$ effect.}
\paragraph{Size of the warp factor.}
The sequestering-violation estimate, Eq.~\eqref{eq:sequestering_violation}, and
the resulting bound on the cutoff, Eq.~\eqref{eq:L_bound}, assume a flat extra
dimension, following Ref.~\cite{Gherghetta:2001sa}: the loop momentum
transmitting SUSY breaking between two branes a proper distance $L$ apart is
then cut off at $\Lambda_{\rm eff}\sim L^{-1}$. In the warped background the
corresponding WKB exponent is $\int_0^{L}dy\,p\,e^{ky}=(p/k)(e^{k\pi R}-1)$, so
that the cutoff is instead
\begin{align}
	\Lambda_{\rm eff} \sim \frac{k}{e^{k\pi R}-1},
	\label{eq:warped_cutoff}
\end{align}
which reduces to $L^{-1}$ in the flat limit $k\to0$ and to the Kaluza--Klein
scale $k\,e^{-k\pi R}$ for $k\pi R\gg1$. Since $Z$ and the matter/messenger
fields sit at opposite ends of the throat, it is this cutoff that controls the
sequestering violation, and requiring it to respect Eq.~\eqref{eq:L_bound} fixes
the size of the warp factor: for $k\sim0.1$, $k\pi R\gtrsim\mathcal{O}(5)$ is
needed at $M_{\rm mess}\sim10^{6}$~GeV and $k\pi R\gtrsim\mathcal{O}(10)$ at
$M_{\rm mess}\sim10^{10}$~GeV, i.e.\ a warp factor
$e^{k\pi R}\gtrsim10^{2}$--$10^{6}$. This leaves the results of
Secs.~\ref{sec:model}--\ref{sec:bino_entropy} intact: the messenger masses and
the soft parameters $M_i$, $B_i$, $A_i$, $m_i^2$ follow from the
UV-brane-localized coupling $h_i(\rho)|X_i|^2$ and the $Z_{3,i}$ symmetry rather
than from loops spanning the throat, so they are independent of
$\Lambda_{\rm eff}$, and the same cutoff also remains above the messenger
scale, $\Lambda_{\rm eff}\gtrsim M_{\rm mess}$, for $k\pi R\lesssim\mathcal{O}(20)$
($k\pi R\lesssim17$ for $M_{\rm mess}\sim10^{10}$~GeV). Depending on $c$,
the contribution transmitted by the bulk hypermultiplet may be further
suppressed by the small overlap of its zero mode with one of the
branes~\cite{Maru:2003mq}.

The radion stabilization can be embedded into our model almost independently, and the predictions presented in the previous sections remain valid.
To suppress the radion abundance, the radion may be strongly coupled to the inflaton~\cite{Linde:1996cx, Nakayama:2011zy, Nakayama:2012mf}.

\section{Conclusion}
We investigate the potential of neutralino dark matter within the framework of gauge-mediated SUSY breaking models with a heavy gravitino of $\mathcal{O}(100)$ TeV. The soft SUSY breaking mass parameters in the MSSM are (almost) exclusively generated by messenger loops, while other possible sources, including anomaly mediation, are suppressed via gravitational SUSY breaking and brane separation. In our setup, both the CP and flavor problems are addressed.

We explore four distinct scenarios for neutralino dark matter: 

(a) Bino-wino coannihilation:
A dedicated collider simulation (Sec.~\ref{sec:model}) shows that the compressed trilepton channel $\chi_2^0\chi_1^\pm\to3\ell+E_T^{\rm miss}$, using a lepton $p_T$ window whose upper edge scales with the compressed mass splitting $\Delta M$, is the most powerful discovery channel for this scenario, with the dilepton channel $\chi_1^+\chi_1^-\to\ell^+\ell^-+E_T^{\rm miss}$ providing a weaker cross-check; our own leading-order simulation is substantially optimistic, since it omits trigger acceptance, pile-up, and several backgrounds present in a full experimental analysis, and we do not quote it as a significance estimate. We instead take the realistic HL-LHC reach from a dedicated analysis of the same topology with an ISR jet and full background estimation~\cite{Duan:2018rls}, which finds $m_{\chi_1^0}\sim330$ GeV at $5\sigma$ and $430$ GeV at $2\sigma$ for $3\,{\rm ab}^{-1}$, consistent with the current CMS Run-2 exclusion of $m_{\chi_2^0}\simeq m_{\chi_1^\pm}\lesssim250$ GeV for $\Delta M\sim30$ GeV. The spin-independent cross section, computed with {\tt micrOMEGAs~6.1.15} for the same benchmark points (Table~\ref{tb:directdetection}), is $\sigma_{\rm SI}\sim10^{-17}$--$10^{-14}$ pb, three to six orders of magnitude below the current LZ bound and also well below typical projections for the neutrino fog; this scenario is therefore essentially inaccessible to direct detection.

(b) Higgsino dark matter:  
In certain regions, the higgsino-like neutralino is the LSP, and the correct abundance is achieved with a mass of $\approx 1.1$ TeV. This scenario marginally avoids current direct detection constraints but is expected to be probed in the near future.

(c) Wino dark matter:  
The wino-like state, with a mass of $\approx 2.8$ TeV, can explain the correct dark matter abundance. As in case (b), the spin-independent cross section tends to be large and is expected to be tested by upcoming experiments. Additionally, this scenario may face tensions with indirect detection experiments, depending on the dark matter density profile and the cosmic propagation model.

(d) Entropy-diluted bino dark matter:
The overproduction of bino-like dark matter can be addressed by an entropy production mechanism involving late-decaying messenger fields. This mechanism dilutes the relic abundance, aligning it with observations while satisfying cosmological constraints such as BBN. The SI cross section is $\mathcal{O}(10^{-12})$ pb, a few percent of the current LZ bound, making this scenario more promising for near-future direct detection than the bino-wino coannihilation scenario of (a).

Future experiments, including HL-LHC searches and direct detection experiments, will be crucial in testing these scenarios and advancing our understanding of neutralino dark matter.

\section*{Acknowledgments}
N.Y. is supported by a start-up grant from Zhejiang University. M.T. is supported by the Fundamental Research Funds for the Central Universities, the One Hundred Talent Program of Sun Yat-sen University, China, and the Guangdong Natural Science Foundation (Project No. 2026A1515012641). J.Z. is supported by Zhejiang University.
We acknowledge the use of Claude Code (Anthropic), with the Claude Opus 5 model, and of the DeepSeek V4.1 model, for literature and reference checking, for assistance in verifying the analytic derivations in Sec.~\ref{sec:radion}, and for editorial assistance in preparing the manuscript. All derivations, numerical analyses, and conclusions were directed and independently verified by the authors, who take full responsibility for the content of this paper.

\appendix

\section{Soft SUSY breaking mass} \label{sec:ap_a}

We derive the formulas for the soft SUSY-breaking mass parameters of the MSSM particles with the following superpotential:
\begin{eqnarray}
	W_{\rm mess} = \lambda_5 X_5 \Psi_5 \bar{\Psi}_5 + \lambda_1 X_1 \Psi_1 \bar{\Psi}_1
	+ \lambda_3 X_3	\Psi_3 \bar{\Psi}_3.
\end{eqnarray}
Messenger loops induce the gaugino masses, sfermion masses and Higgs masses. They can be extracted by wave function renormalization constants using the analytic continuation method~\cite{Giudice:1997ni,Arkani-Hamed:1998mzz}.
The generated gaugino masses are
\begin{eqnarray}
	M_{\tilde{b}} &\simeq& \frac{g_1^2}{16\pi^2} (B_5 + B_1)   ,
	\nonumber \\
	M_{\tilde{w}} &\simeq& \frac{g_2^2}{16\pi^2}
	(B_5),
	\nonumber \\
	M_{\tilde{g}} &\simeq&
	\frac{g_3^2}{16\pi^2} (B_5 + B_3).
\end{eqnarray}
When $x_i \equiv |B_i/M_{\rm mess}|$ is close one, $B_i$ should be replaced to $B_i \mathcal{G}(x_i)$, where~\cite{Martin:1996zb}
\begin{eqnarray}
	\mathcal{G}(x)=\frac{1}{x^2} \left[(1+x) \ln (1+x) \right] + (x \to -x).
\end{eqnarray}

The sfermion masses are
\begin{eqnarray}
	m_{\tilde Q}^2 &\simeq& \frac{1}{256 \pi^4} \left[\frac{8}{3}g_3^4 (B_5^2 + B_3^2)
	+ \frac{3}{2}g_2^4 B_5^2  + \frac{1}{30}  g_1^4 (B_5^2 + B_1^2) \right],
\nonumber \\ 
m_{\tilde U}^2 &\simeq& \frac{1}{256 \pi^4}\left[\frac{8}{3}g_3^4 (B_5^2 + B_3^2)
 + \frac{8}{15}  g_1^4 (B_5^2 + B_1^2) \right],
\nonumber \\ 
m_{\tilde D}^2 &\simeq&	\frac{1}{256 \pi^4}	\left[\frac{8}{3}g_3^4 (B_5^2 + B_3^2) 
+ \frac{2}{15}  g_1^4 (B_5^2 + B_1^2)
\right],
\nonumber \\ 
m_{\tilde L}^2 &\simeq&	\frac{1}{256 \pi^4}	\left[\frac{3}{2}g_2^4 B_5^2 + \frac{3}{10}  g_1^4 (B_5^2 + B_1^2) \right],
\nonumber \\ 
m_{\tilde E}^2 &\simeq& \frac{1}{256 \pi^4}\left[ \frac{6}{5} g_1^4 (B_5^2 + B_1^2) \right],
\nonumber \\ 
m_{H_u}^2 &=& m_{H_d}^2 = m_{\tilde L}^2	.	
\end{eqnarray}
For $x_i$ close to one, $B_i^2$ should be replaced to $B_i^2 \mathcal{F}(x_i)$, where
\begin{eqnarray}
	\mathcal{F}(x)=\frac{1+x}{x^2}
	\left[
	\ln(1+x)- 2 {\rm Li}_2(x/(1+x)) + \frac{1}{2}{\rm Li}_2(2x/(1+x))
	\right] + (x \to -x) .
\end{eqnarray}

\bibliographystyle{utphys}
\bibliography{refs}

@article{Alwall:2014hca,
    author = "Alwall, J. and Frederix, R. and Frixione, S. and Hirschi, V. and Maltoni, F. and Mattelaer, O. and Shao, H.-S. and Stelzer, T. and Torrielli, P. and Zaro, M.",
    title = "{The automated computation of tree-level and next-to-leading order differential cross sections, and their matching to parton shower simulations}",
    eprint = "1405.0301",
    archivePrefix = "arXiv",
    primaryClass = "hep-ph",
    doi = "10.1007/JHEP07(2014)079",
    journal = "JHEP",
    volume = "07",
    pages = "079",
    year = "2014"
}

@article{Sjostrand:2014zea,
    author = "Sjostrand, Torbjorn and Ask, Stefan and Christiansen, Jesper R. and Corke, Richard and Desai, Nishita and Ilten, Philip and Mrenna, Stephen and Prestel, Stefan and Rasmussen, Christine O. and Skands, Peter Z.",
    title = "{An introduction to PYTHIA 8.2}",
    eprint = "1410.3012",
    archivePrefix = "arXiv",
    primaryClass = "hep-ph",
    doi = "10.1016/j.cpc.2015.01.024",
    journal = "Comput. Phys. Commun.",
    volume = "191",
    pages = "159--177",
    year = "2015"
}

@article{deFavereau:2013fsa,
    author = "de Favereau, J. and Delaere, C. and Demin, P. and Giammanco, A. and Lemaitre, V. and Mertens, A. and Selvaggi, M.",
    collaboration = "DELPHES 3",
    title = "{DELPHES 3, A modular framework for fast simulation of a generic collider experiment}",
    eprint = "1307.6346",
    archivePrefix = "arXiv",
    primaryClass = "hep-ex",
    doi = "10.1007/JHEP02(2014)057",
    journal = "JHEP",
    volume = "02",
    pages = "057",
    year = "2014"
}

@article{Moroi:1993mb,
    author = "Moroi, T. and Murayama, H. and Yamaguchi, Masahiro",
    title = "{Cosmological constraints on the light stable gravitino}",
    reportNumber = "TU-424",
    doi = "10.1016/0370-2693(93)91434-O",
    journal = "Phys. Lett. B",
    volume = "303",
    pages = "289--294",
    year = "1993"
}

@article{Bolz:2000fu,
    author = "Bolz, M. and Brandenburg, A. and Buchmuller, W.",
    title = "{Thermal production of gravitinos}",
    eprint = "hep-ph/0012052",
    archivePrefix = "arXiv",
    reportNumber = "DESY-00-167",
    doi = "10.1016/S0550-3213(01)00132-8",
    journal = "Nucl. Phys. B",
    volume = "606",
    pages = "518--544",
    year = "2001",
    note = "[Erratum: Nucl.Phys.B 790, 336--337 (2008)]"
}

@article{Kawasaki:2008qe,
    author = "Kawasaki, Masahiro and Kohri, Kazunori and Moroi, Takeo and Yotsuyanagi, Akira",
    title = "{Big-Bang Nucleosynthesis and Gravitino}",
    eprint = "0804.3745",
    archivePrefix = "arXiv",
    primaryClass = "hep-ph",
    reportNumber = "TU-812",
    doi = "10.1103/PhysRevD.78.065011",
    journal = "Phys. Rev. D",
    volume = "78",
    pages = "065011",
    year = "2008"
}

@article{Eberl:2020fml,
    author = "Eberl, Helmut and Gialamas, Ioannis D. and Spanos, Vassilis C.",
    title = "{Gravitino thermal production revisited}",
    eprint = "2010.14621",
    archivePrefix = "arXiv",
    primaryClass = "hep-ph",
    doi = "10.1103/PhysRevD.103.075025",
    journal = "Phys. Rev. D",
    volume = "103",
    number = "7",
    pages = "075025",
    year = "2021"
}

@article{Eberl:2024pxr,
    author = "Eberl, Helmut and Gialamas, Ioannis D. and Spanos, Vassilis C.",
    title = "{Gravitino thermal production, dark matter, and~reheating of the Universe}",
    eprint = "2408.16043",
    archivePrefix = "arXiv",
    primaryClass = "hep-ph",
    reportNumber = "FTPI-MINN-24-20",
    doi = "10.1088/1475-7516/2025/01/079",
    journal = "JCAP",
    volume = "01",
    pages = "079",
    year = "2025"
}

@article{Izawa:2010ym,
    author = "Izawa, K. -I. and Kugo, T. and Yanagida, T. T.",
    title = "{Gravitational Supersymmetry Breaking}",
    eprint = "1008.4641",
    archivePrefix = "arXiv",
    primaryClass = "hep-ph",
    reportNumber = "YITP-10-73, IPMU-10-0143",
    doi = "10.1143/PTP.125.261",
    journal = "Prog. Theor. Phys.",
    volume = "125",
    pages = "261--264",
    year = "2011"
}

@article{Harigaya:2017akm,
    author = "Harigaya, Keisuke and Yanagida, Tsutomu T. and Yokozaki, Norimi",
    title = "{Axion induced SUSY breaking and focus point gaugino mediation}",
    eprint = "1710.02204",
    archivePrefix = "arXiv",
    primaryClass = "hep-ph",
    reportNumber = "IPMU17-0134, TU-1047",
    doi = "10.1016/j.physletb.2022.137386",
    journal = "Phys. Lett. B",
    volume = "833",
    pages = "137386",
    year = "2022"
}

@article{Ciuchini:1998ix,
    author = "Ciuchini, Marco and others",
    title = "{Delta M(K) and epsilon(K) in SUSY at the next-to-leading order}",
    eprint = "hep-ph/9808328",
    archivePrefix = "arXiv",
    reportNumber = "TUM-HEP-320-98, ROME1-1217-98, EDINBURGH-98-14, ROM2F-98-33",
    doi = "10.1088/1126-6708/1998/10/008",
    journal = "JHEP",
    volume = "10",
    pages = "008",
    year = "1998"
}

@article{Asano:2015kvj,
    author = "Asano, Masaki and Nakai, Yuichiro and Yokozaki, Norimi",
    title = "{Low-Scale Gauge Mediation with a 100 TeV Gravitino}",
    eprint = "1512.02201",
    archivePrefix = "arXiv",
    primaryClass = "hep-ph",
    doi = "10.1103/PhysRevD.93.055023",
    journal = "Phys. Rev. D",
    volume = "93",
    number = "5",
    pages = "055023",
    year = "2016"
}

@article{Asano:2016oik,
    author = "Asano, Masaki and Yokozaki, Norimi",
    title = "{Light Higgsino as the tail of the $\mu$-$B_\mu$ solution}",
    eprint = "1601.00652",
    archivePrefix = "arXiv",
    primaryClass = "hep-ph",
    doi = "10.1103/PhysRevD.93.095002",
    journal = "Phys. Rev. D",
    volume = "93",
    number = "9",
    pages = "095002",
    year = "2016"
}

@article{Nagai:2017tuf,
    author = "Nagai, Ryo and Yokozaki, Norimi",
    title = "{Light Higgsino and Gluino in $R$-invariant Direct Gauge Mediation}",
    eprint = "1711.04431",
    archivePrefix = "arXiv",
    primaryClass = "hep-ph",
    reportNumber = "TU-1051",
    doi = "10.1016/j.physletb.2018.01.025",
    journal = "Phys. Lett. B",
    volume = "778",
    pages = "309--315",
    year = "2018"
}

@article{Fujii:2002fv,
    author = "Fujii, Masaaki and Yanagida, T.",
    title = "{Natural gravitino dark matter and thermal leptogenesis in gauge mediated supersymmetry breaking models}",
    eprint = "hep-ph/0208191",
    archivePrefix = "arXiv",
    reportNumber = "CERN-TH-2002-200, UT-02-48",
    doi = "10.1016/S0370-2693(02)02958-1",
    journal = "Phys. Lett. B",
    volume = "549",
    pages = "273--283",
    year = "2002"
}

@article{Hamaguchi:2014sea,
    author = "Hamaguchi, Koichi and Ibe, Masahiro and Yanagida, Tsutomu T. and Yokozaki, Norimi",
    title = "{Testing the Minimal Direct Gauge Mediation at the LHC}",
    eprint = "1403.1398",
    archivePrefix = "arXiv",
    primaryClass = "hep-ph",
    doi = "10.1103/PhysRevD.90.015027",
    journal = "Phys. Rev. D",
    volume = "90",
    number = "1",
    pages = "015027",
    year = "2014"
}

@article{Dimopoulos:1996gy,
    author = "Dimopoulos, S. and Giudice, G. F. and Pomarol, A.",
    title = "{Dark matter in theories of gauge mediated supersymmetry breaking}",
    eprint = "hep-ph/9607225",
    archivePrefix = "arXiv",
    reportNumber = "CERN-TH-96-171",
    doi = "10.1016/S0370-2693(96)01241-5",
    journal = "Phys. Lett. B",
    volume = "389",
    pages = "37--42",
    year = "1996"
}

@article{Han:1997wn,
    author = "Han, Tao and Hempfling, Ralf",
    title = "{Messenger sneutrinos as cold dark matter}",
    eprint = "hep-ph/9708264",
    archivePrefix = "arXiv",
    reportNumber = "UCD-97-16",
    doi = "10.1016/S0370-2693(97)01205-7",
    journal = "Phys. Lett. B",
    volume = "415",
    pages = "161--169",
    year = "1997"
}

@article{Martin:1996zb,
    author = "Martin, Stephen P.",
    title = "{Generalized messengers of supersymmetry breaking and the sparticle mass spectrum}",
    eprint = "hep-ph/9608224",
    archivePrefix = "arXiv",
    doi = "10.1103/PhysRevD.55.3177",
    journal = "Phys. Rev. D",
    volume = "55",
    pages = "3177--3187",
    year = "1997"
}

@mastersthesis{Moroi:1995fs,
	author = "Moroi, Takeo",
	title = "{Effects of the gravitino on the inflationary universe}",
	eprint = "hep-ph/9503210",
	archivePrefix = "arXiv",
	reportNumber = "TU-479",
	school = "Tohoku University",
	type = "Other thesis",
	month = "3",
	year = "1995"
}

@article{Giudice:1997ni,
    author = "Giudice, G. F. and Rattazzi, R.",
    title = "{Extracting supersymmetry breaking effects from wave function renormalization}",
    eprint = "hep-ph/9706540",
    archivePrefix = "arXiv",
    reportNumber = "CERN-TH-97-145",
    doi = "10.1016/S0550-3213(97)00647-0",
    journal = "Nucl. Phys. B",
    volume = "511",
    pages = "25--44",
    year = "1998"
}

@article{Arkani-Hamed:1998mzz,
    author = "Arkani-Hamed, Nima and Giudice, Gian F. and Luty, Markus A. and Rattazzi, Riccardo",
    title = "{Supersymmetry breaking loops from analytic continuation into superspace}",
    eprint = "hep-ph/9803290",
    archivePrefix = "arXiv",
    reportNumber = "SLAC-PUB-7764, CERN-TH-98-47, UMDHEP-98-47",
    doi = "10.1103/PhysRevD.58.115005",
    journal = "Phys. Rev. D",
    volume = "58",
    pages = "115005",
    year = "1998"
}

@article{Dine:1981za,
    author = "Dine, Michael and Fischler, Willy and Srednicki, Mark",
    editor = "Leveille, J. P. and Sulak, L. R. and Unger, D. G.",
    title = "{Supersymmetric Technicolor}",
    reportNumber = "PRINT-81-0278 (IAS,PRINCETON)",
    doi = "10.1016/0550-3213(81)90582-4",
    journal = "Nucl. Phys. B",
    volume = "189",
    pages = "575--593",
    year = "1981"
}

@article{Dimopoulos:1981au,
    author = "Dimopoulos, Savas and Raby, Stuart",
    title = "{Supercolor}",
    reportNumber = "SLAC-PUB-2719, Print-81-0215 (ITP,STANFORD), NSF-ITP-81-24",
    doi = "10.1016/0550-3213(81)90430-2",
    journal = "Nucl. Phys. B",
    volume = "192",
    pages = "353--368",
    year = "1981"
}

@article{Dine:1981gu,
    author = "Dine, Michael and Fischler, Willy",
    title = "{A Phenomenological Model of Particle Physics Based on Supersymmetry}",
    reportNumber = "PRINT-82-0029 (IAS,PRINCETON)",
    doi = "10.1016/0370-2693(82)91241-2",
    journal = "Phys. Lett. B",
    volume = "110",
    pages = "227--231",
    year = "1982"
}

@article{Nappi:1982hm,
    author = "Nappi, Chiara R. and Ovrut, Burt A.",
    title = "{Supersymmetric Extension of the SU(3) x SU(2) x U(1) Model}",
    reportNumber = "PRINT-82-0093 (IAS,PRINCETON)",
    doi = "10.1016/0370-2693(82)90418-X",
    journal = "Phys. Lett. B",
    volume = "113",
    pages = "175--179",
    year = "1982"
}

@article{Gabrielli:1997jp,
    author = "Gabrielli, Emidio and Sarid, Uri",
    title = "{Low-energy signals for a minimal gauge mediated model}",
    eprint = "hep-ph/9707546",
    archivePrefix = "arXiv",
    reportNumber = "UND-HEP-97-US02",
    doi = "10.1103/PhysRevLett.79.4752",
    journal = "Phys. Rev. Lett.",
    volume = "79",
    pages = "4752--4755",
    year = "1997"
}

@article{Dine:1993yw,
    author = "Dine, Michael and Nelson, Ann E.",
    title = "{Dynamical supersymmetry breaking at low-energies}",
    eprint = "hep-ph/9303230",
    archivePrefix = "arXiv",
    reportNumber = "SCIPP-93-03, UCSD-PTH-93-05",
    doi = "10.1103/PhysRevD.48.1277",
    journal = "Phys. Rev. D",
    volume = "48",
    pages = "1277--1287",
    year = "1993"
}

@article{Dine:1994vc,
    author = "Dine, Michael and Nelson, Ann E. and Shirman, Yuri",
    title = "{Low-energy dynamical supersymmetry breaking simplified}",
    eprint = "hep-ph/9408384",
    archivePrefix = "arXiv",
    reportNumber = "SCIPP-94-21, UW-PT-94-07",
    doi = "10.1103/PhysRevD.51.1362",
    journal = "Phys. Rev. D",
    volume = "51",
    pages = "1362--1370",
    year = "1995"
}

@article{Dine:1995ag,
    author = "Dine, Michael and Nelson, Ann E. and Nir, Yosef and Shirman, Yuri",
    title = "{New tools for low-energy dynamical supersymmetry breaking}",
    eprint = "hep-ph/9507378",
    archivePrefix = "arXiv",
    reportNumber = "SCIPP-95-32, UW-PT-95-08, WIS-95-29-JUL-PH",
    doi = "10.1103/PhysRevD.53.2658",
    journal = "Phys. Rev. D",
    volume = "53",
    pages = "2658--2669",
    year = "1996"
}

@article{Evans:2010ru,
    author = "Evans, Jason L. and Sudano, Matthew and Yanagida, Tsutomu T.",
    title = "{A CP-safe solution of the $\mu$/B$\mu$ problem of gauge mediation}",
    eprint = "1008.3165",
    archivePrefix = "arXiv",
    primaryClass = "hep-ph",
    reportNumber = "IPMU10-0139",
    doi = "10.1016/j.physletb.2010.10.013",
    journal = "Phys. Lett. B",
    volume = "696",
    pages = "348--351",
    year = "2011"
}

@article{Moroi:2011fi,
    author = "Moroi, Takeo and Yokozaki, Norimi",
    title = "{SUSY CP Problem in Gauge Mediation Model}",
    eprint = "1105.3294",
    archivePrefix = "arXiv",
    primaryClass = "hep-ph",
    doi = "10.1016/j.physletb.2011.06.020",
    journal = "Phys. Lett. B",
    volume = "701",
    pages = "568--575",
    year = "2011"
}

@article{Evans:2022oho,
    author = "Evans, Jason L. and Yanagida, Tsutomu T. and Yokozaki, Norimi",
    title = "{Flavor- and CP-safe explanation of g$_{\mu}$ \ensuremath{-} 2 anomaly}",
    eprint = "2205.14906",
    archivePrefix = "arXiv",
    primaryClass = "hep-ph",
    doi = "10.1007/JHEP03(2023)024",
    journal = "JHEP",
    volume = "03",
    pages = "024",
    year = "2023"
}

@article{Rattazzi:1996fb,
    author = "Rattazzi, Riccardo and Sarid, Uri",
    title = "{Large tan Beta in gauge mediated SUSY breaking models}",
    eprint = "hep-ph/9612464",
    archivePrefix = "arXiv",
    reportNumber = "UND-HEP-96-US01, CERN-TH-96-349",
    doi = "10.1016/S0550-3213(97)00363-5",
    journal = "Nucl. Phys. B",
    volume = "501",
    pages = "297--331",
    year = "1997"
}

@article{OHare:2021utq,
    author = "O'Hare, Ciaran A. J.",
    title = "{New Definition of the Neutrino Floor for Direct Dark Matter Searches}",
    eprint = "2109.03116",
    archivePrefix = "arXiv",
    primaryClass = "hep-ph",
    doi = "10.1103/PhysRevLett.127.251802",
    journal = "Phys. Rev. Lett.",
    volume = "127",
    number = "25",
    pages = "251802",
    year = "2021"
}

@article{Belanger:2001fz,
    author = "Belanger, G. and Boudjema, F. and Pukhov, A. and Semenov, A.",
    title = "{MicrOMEGAs: A Program for calculating the relic density in the MSSM}",
    eprint = "hep-ph/0112278",
    archivePrefix = "arXiv",
    reportNumber = "LAPTH-881-01",
    doi = "10.1016/S0010-4655(02)00596-9",
    journal = "Comput. Phys. Commun.",
    volume = "149",
    pages = "103--120",
    year = "2002"
}

@article{Belanger:2004yn,
    author = "Belanger, G. and Boudjema, F. and Pukhov, A. and Semenov, A.",
    title = "{micrOMEGAs: Version 1.3}",
    eprint = "hep-ph/0405253",
    archivePrefix = "arXiv",
    reportNumber = "LAPTH-1044",
    doi = "10.1016/j.cpc.2005.12.005",
    journal = "Comput. Phys. Commun.",
    volume = "174",
    pages = "577--604",
    year = "2006"
}

@article{LZ:2024zvo,
    author = "Aalbers, J. and others",
    collaboration = "LZ",
    title = "{Dark Matter Search Results from 4.2 Tonne-Years of Exposure of the LUX-ZEPLIN (LZ) Experiment}",
    eprint = "2410.17036",
    archivePrefix = "arXiv",
    primaryClass = "hep-ex",
    reportNumber = "FERMILAB-PUB-24-0796-V",
    month = "10",
    year = "2024"
}

@article{Bottaro:2021snn,
    author = "Bottaro, Salvatore and Buttazzo, Dario and Costa, Marco and Franceschini, Roberto and Panci, Paolo and Redigolo, Diego and Vittorio, Ludovico",
    title = "{Closing the window on WIMP Dark Matter}",
    eprint = "2107.09688",
    archivePrefix = "arXiv",
    primaryClass = "hep-ph",
    doi = "10.1140/epjc/s10052-021-09917-9",
    journal = "Eur. Phys. J. C",
    volume = "82",
    number = "1",
    pages = "31",
    year = "2022"
}

@article{Bottaro:2022one,
    author = "Bottaro, Salvatore and Buttazzo, Dario and Costa, Marco and Franceschini, Roberto and Panci, Paolo and Redigolo, Diego and Vittorio, Ludovico",
    title = "{The last complex WIMPs standing}",
    eprint = "2205.04486",
    archivePrefix = "arXiv",
    primaryClass = "hep-ph",
    reportNumber = "CERN-TH-2022-080",
    doi = "10.1140/epjc/s10052-022-10918-5",
    journal = "Eur. Phys. J. C",
    volume = "82",
    number = "11",
    pages = "992",
    year = "2022"
}

@article{Hisano:2006nn,
    author = "Hisano, Junji and Matsumoto, Shigeki and Nagai, Minoru and Saito, Osamu and Senami, Masato",
    title = "{Non-perturbative effect on thermal relic abundance of dark matter}",
    eprint = "hep-ph/0610249",
    archivePrefix = "arXiv",
    reportNumber = "KEK-TH-1111",
    doi = "10.1016/j.physletb.2007.01.012",
    journal = "Phys. Lett. B",
    volume = "646",
    pages = "34--38",
    year = "2007"
}

@article{Hryczuk:2010zi,
    author = "Hryczuk, Andrzej and Iengo, Roberto and Ullio, Piero",
    title = "{Relic densities including Sommerfeld enhancements in the MSSM}",
    eprint = "1010.2172",
    archivePrefix = "arXiv",
    primaryClass = "hep-ph",
    doi = "10.1007/JHEP03(2011)069",
    journal = "JHEP",
    volume = "03",
    pages = "069",
    year = "2011"
}

@article{Hryczuk:2011vi,
    author = "Hryczuk, Andrzej and Iengo, Roberto",
    title = "{The one-loop and Sommerfeld electroweak corrections to the Wino dark matter annihilation}",
    eprint = "1111.2916",
    archivePrefix = "arXiv",
    primaryClass = "hep-ph",
    doi = "10.1007/JHEP01(2012)163",
    journal = "JHEP",
    volume = "01",
    pages = "163",
    year = "2012",
    note = "[Erratum: JHEP 06, 137 (2012)]"
}

@article{Boveia:2022adi,
    author = "Boveia, Antonio and others",
    title = "{Snowmass 2021 Dark Matter Complementarity Report}",
    eprint = "2211.07027",
    archivePrefix = "arXiv",
    primaryClass = "hep-ex",
    month = "11",
    year = "2022"
}

@article{Gherghetta:2001sa,
    author = "Gherghetta, Tony and Riotto, Antonio",
    title = "{Gravity mediated supersymmetry breaking in the brane world}",
    eprint = "hep-th/0110022",
    archivePrefix = "arXiv",
    reportNumber = "UNIL-IPT-01-15",
    doi = "10.1016/S0550-3213(01)00637-X",
    journal = "Nucl. Phys. B",
    volume = "623",
    pages = "97--125",
    year = "2002"
}

@article{Harigaya:2014dwa,
    author = "Harigaya, Keisuke and Kaneta, Kunio and Matsumoto, Shigeki",
    title = "{Gaugino coannihilations}",
    eprint = "1403.0715",
    archivePrefix = "arXiv",
    primaryClass = "hep-ph",
    reportNumber = "IPMU-14-0047",
    doi = "10.1103/PhysRevD.89.115021",
    journal = "Phys. Rev. D",
    volume = "89",
    number = "11",
    pages = "115021",
    year = "2014"
}

@article{ATLAS:2024lda,
    author = "Aad, Georges and others",
    collaboration = "ATLAS",
    title = "{The quest to discover supersymmetry at the ATLAS experiment}",
    eprint = "2403.02455",
    archivePrefix = "arXiv",
    primaryClass = "hep-ex",
    reportNumber = "CERN-EP-2024-056",
    month = "3",
    year = "2024"
}

@article{ATLAS:2024vnc,
    author = "Aad, Georges and others",
    collaboration = "ATLAS",
    title = "{Search for displaced leptons in $\sqrt{s}=13$ TeV and $13.6$ TeV $pp$ collisions with the ATLAS detector}",
    eprint = "2410.16835",
    archivePrefix = "arXiv",
    primaryClass = "hep-ex",
    reportNumber = "CERN-EP-2024-257",
    month = "10",
    year = "2024"
}

@article{ATLAS:2020zms,
    author = "Aad, Georges and others",
    collaboration = "ATLAS",
    title = "{Search for heavy Higgs bosons decaying into two tau leptons with the ATLAS detector using $pp$ collisions at $\sqrt{s}=13$ TeV}",
    eprint = "2002.12223",
    archivePrefix = "arXiv",
    primaryClass = "hep-ex",
    reportNumber = "CERN-EP-2020-014",
    doi = "10.1103/PhysRevLett.125.051801",
    journal = "Phys. Rev. Lett.",
    volume = "125",
    number = "5",
    pages = "051801",
    year = "2020"
}

@article{Choi:2020wdq,
    author = "Choi, Gongjun and Yanagida, Tsutomu T. and Yokozaki, Norimi",
    title = "{The Upper Bound of the Second Higgs Boson Mass in Minimal Gauge Mediation with the Gravitino Warm Dark Matter}",
    eprint = "2012.03266",
    archivePrefix = "arXiv",
    primaryClass = "hep-ph",
    doi = "10.1007/JHEP04(2021)024",
    journal = "JHEP",
    volume = "04",
    pages = "024",
    year = "2021"
}

@article{Yanagida:1994vq,
    author = "Yanagida, T.",
    title = "{Naturally light Higgs doublets in the supersymmetric grand unified theories with dynamical symmetry breaking}",
    eprint = "hep-ph/9409329",
    archivePrefix = "arXiv",
    reportNumber = "TU-464",
    doi = "10.1016/0370-2693(94)01500-C",
    journal = "Phys. Lett. B",
    volume = "344",
    pages = "211--216",
    year = "1995"
}

@article{Hotta:1995cd,
    author = "Hotta, T. and Izawa, K. I. and Yanagida, T.",
    title = "{Dynamical models for light Higgs doublets in supersymmetric grand unified theories}",
    eprint = "hep-ph/9509201",
    archivePrefix = "arXiv",
    reportNumber = "UT-717",
    doi = "10.1103/PhysRevD.53.3913",
    journal = "Phys. Rev. D",
    volume = "53",
    pages = "3913--3919",
    year = "1996"
}

@article{Linde:1996cx,
    author = "Linde, Andrei D.",
    title = "{Relaxing the cosmological moduli problem}",
    eprint = "hep-th/9601083",
    archivePrefix = "arXiv",
    reportNumber = "SU-ITP-96-03",
    doi = "10.1103/PhysRevD.53.R4129",
    journal = "Phys. Rev. D",
    volume = "53",
    pages = "R4129--R4132",
    year = "1996"
}

@article{Nakayama:2011zy,
    author = "Nakayama, Kazunori and Takahashi, Fuminobu and Yanagida, Tsutomu T.",
    title = "{Cosmological Moduli Problem in Low Cutoff Theory}",
    eprint = "1112.0418",
    archivePrefix = "arXiv",
    primaryClass = "hep-ph",
    reportNumber = "UT-11-41, TU-896",
    doi = "10.1103/PhysRevD.86.043507",
    journal = "Phys. Rev. D",
    volume = "86",
    pages = "043507",
    year = "2012"
}

@article{Nakayama:2012mf,
    author = "Nakayama, Kazunori and Takahashi, Fuminobu and Yanagida, Tsutomu T.",
    title = "{Gravity mediation without a Polonyi problem}",
    eprint = "1203.2085",
    archivePrefix = "arXiv",
    primaryClass = "hep-ph",
    reportNumber = "UT-12-06, TU-901, IPMU12-0027",
    doi = "10.1016/j.physletb.2012.06.072",
    journal = "Phys. Lett. B",
    volume = "714",
    pages = "256--261",
    year = "2012"
}

@article{Nomura:2001ub,
    author = "Nomura, Yasunori and Suzuki, Koshiro",
    title = "{Gauge mediation models with neutralino dark matter}",
    eprint = "hep-ph/0110040",
    archivePrefix = "arXiv",
    reportNumber = "UCB-PTH-00-45, LBNL-47230, UT-955",
    doi = "10.1103/PhysRevD.68.075005",
    journal = "Phys. Rev. D",
    volume = "68",
    pages = "075005",
    year = "2003"
}

@article{Roussy:2022cmp,
    author = "Roussy, Tanya S. and others",
    title = "{An improved bound on the electron\textquoteright{}s electric dipole moment}",
    eprint = "2212.11841",
    archivePrefix = "arXiv",
    primaryClass = "physics.atom-ph",
    doi = "10.1126/science.adg4084",
    journal = "Science",
    volume = "381",
    number = "6653",
    pages = "adg4084",
    year = "2023"
}

@article{Moroi:2013sfa,
    author = "Moroi, Takeo and Nagai, Minoru",
    title = "{Probing Supersymmetric Model with Heavy Sfermions Using Leptonic Flavor and CP Violations}",
    eprint = "1303.0668",
    archivePrefix = "arXiv",
    primaryClass = "hep-ph",
    reportNumber = "UT-13-08",
    doi = "10.1016/j.physletb.2013.04.049",
    journal = "Phys. Lett. B",
    volume = "723",
    pages = "107--112",
    year = "2013"
}

@article{Ellis:1990nz,
    author = "Ellis, John R. and Ridolfi, Giovanni and Zwirner, Fabio",
    title = "{Radiative corrections to the masses of supersymmetric Higgs bosons}",
    reportNumber = "CERN-TH-5946-90, GEF-TH-26-1990",
    doi = "10.1016/0370-2693(91)90863-L",
    journal = "Phys. Lett. B",
    volume = "257",
    pages = "83--91",
    year = "1991"
}

@article{Okada:1990vk,
    author = "Okada, Yasuhiro and Yamaguchi, Masahiro and Yanagida, Tsutomu",
    title = "{Upper bound of the lightest Higgs boson mass in the minimal supersymmetric standard model}",
    reportNumber = "TU-360-REV, TU-360",
    doi = "10.1143/ptp/85.1.1",
    journal = "Prog. Theor. Phys.",
    volume = "85",
    pages = "1--6",
    year = "1991"
}

@article{Haber:1990aw,
    author = "Haber, Howard E. and Hempfling, Ralf",
    title = "{Can the mass of the lightest Higgs boson of the minimal supersymmetric model be larger than m(Z)?}",
    reportNumber = "SCIPP-90-42",
    doi = "10.1103/PhysRevLett.66.1815",
    journal = "Phys. Rev. Lett.",
    volume = "66",
    pages = "1815--1818",
    year = "1991"
}

@article{Okada:1990gg,
    author = "Okada, Y. and Yamaguchi, Masahiro and Yanagida, T.",
    title = "{Renormalization group analysis on the Higgs mass in the softly broken supersymmetric standard model}",
    reportNumber = "TU-363",
    doi = "10.1016/0370-2693(91)90642-4",
    journal = "Phys. Lett. B",
    volume = "262",
    pages = "54--58",
    year = "1991"
}

@article{Slavich:2020zjv,
    author = "Slavich, P. and others",
    editor = "Slavich, P. and Heinemeyer, S.",
    title = "{Higgs-mass predictions in the MSSM and beyond}",
    eprint = "2012.15629",
    archivePrefix = "arXiv",
    primaryClass = "hep-ph",
    reportNumber = "DESY 20-229, DESY-20-229, IFT-UAM/CSIC-20-184, FR-PHENO-2020-021, KA-TP-23-2020, MPP-2020-235, KA-TP-23-2020,
  MPP-2020-235, P3H-20-086, TTK-20-53, FERMILAB-PUB-21-575-T",
    doi = "10.1140/epjc/s10052-021-09198-2",
    journal = "Eur. Phys. J. C",
    volume = "81",
    number = "5",
    pages = "450",
    year = "2021"
}

@article{MEGII:2023ltw,
    author = "Afanaciev, K. and others",
    collaboration = "MEG II",
    title = "{A search for $\mu ^+ \rightarrow e^+ \gamma $ with the first dataset of the MEG~II experiment}",
    eprint = "2310.12614",
    archivePrefix = "arXiv",
    primaryClass = "hep-ex",
    doi = "10.1140/epjc/s10052-024-12416-2",
    journal = "Eur. Phys. J. C",
    volume = "84",
    number = "3",
    pages = "216",
    year = "2024",
    note = "[Erratum: Eur.Phys.J.C 84, 1042 (2024)]"
}

@article{Goldberger:1999uk,
  author         = "Goldberger, Walter D. and Wise, Mark B.",
  title          = "{Modulus stabilization with bulk fields}",
  journal        = "Phys. Rev. Lett.",
  volume         = "83",
  year           = "1999",
  pages          = "4922--4925",
  doi            = "10.1103/PhysRevLett.83.4922",
  eprint         = "hep-ph/9907447",
  archivePrefix  = "arXiv"
}

@article{Randall:1999ee,
  author         = "Randall, Lisa and Sundrum, Raman",
  title          = "{A Large mass hierarchy from a small extra dimension}",
  journal        = "Phys. Rev. Lett.",
  volume         = "83",
  year           = "1999",
  pages          = "3370--3373",
  doi            = "10.1103/PhysRevLett.83.3370",
  eprint         = "hep-ph/9905221",
  archivePrefix  = "arXiv"
}

@article{Maru:2003mq,
  author         = "Maru, Nobuhito and Okada, Nobuchika",
  title          = "{Supersymmetric radius stabilization in warped extra dimensions}",
  journal        = "Phys. Rev. D",
  volume         = "70",
  year           = "2004",
  pages          = "025002",
  doi            = "10.1103/PhysRevD.70.025002",
  eprint         = "hep-th/0312148",
  archivePrefix  = "arXiv"
}

@article{Eto:2004yk,
  author         = "Eto, Minoru and Maru, Nobuhito and Sakai, Norisuke",
  title          = "{Radius stabilization in a supersymmetric warped compactification}",
  journal        = "Phys. Rev. D",
  volume         = "70",
  year           = "2004",
  pages          = "086002",
  doi            = "10.1103/PhysRevD.70.086002",
  eprint         = "hep-th/0403009",
  archivePrefix  = "arXiv"
}

@article{PaccettiCorreia:2006rpo,
  author         = "Paccetti Correia, Filipe and Schmidt, Michael G. and Tavartkiladze, Zurab",
  title          = "{4D superfield reduction of 5D orbifold SUGRA and heterotic M-theory}",
  journal        = "Nucl. Phys. B",
  volume         = "751",
  year           = "2006",
  pages          = "222--259",
  doi            = "10.1016/j.nuclphysb.2006.06.007",
  eprint         = "hep-th/0602173",
  archivePrefix  = "arXiv"
}

@article{Marti:2001iw,
  author         = "Marti, Daniel and Pomarol, Alex",
  title          = "{Supersymmetric theories with compact extra dimensions in N=1 superfields}",
  journal        = "Phys. Rev. D",
  volume         = "64",
  year           = "2001",
  pages          = "105025",
  doi            = "10.1103/PhysRevD.64.105025",
  eprint         = "hep-th/0106256",
  archivePrefix  = "arXiv"
}

@article{PaccettiCorreia:2004suu,
  author         = "Paccetti Correia, Filipe and Schmidt, Michael G. and Tavartkiladze, Zurab",
  title          = "{Superfield approach to 5D conformal SUGRA and the radion}",
  journal        = "Nucl. Phys. B",
  volume         = "709",
  year           = "2005",
  pages          = "141--170",
  doi            = "10.1016/j.nuclphysb.2004.12.005",
  eprint         = "hep-th/0408138",
  archivePrefix  = "arXiv"
}

@article{PICO:2019vsc,
  author         = "Amole, C. and others",
  collaboration  = "PICO",
  title          = "{Dark Matter Search Results from the Complete Exposure of the PICO-60 C$_3$F$_8$ Bubble Chamber}",
  journal        = "Phys. Rev. D",
  volume         = "100",
  year           = "2019",
  number         = "2",
  pages          = "022001",
  doi            = "10.1103/PhysRevD.100.022001",
  eprint         = "1902.04031",
  archivePrefix  = "arXiv",
  primaryClass   = "astro-ph.CO"
}

@article{CMS:2021edw,
  author         = "Tumasyan, Armen and others",
  collaboration  = "CMS",
  title          = "{Search for supersymmetry in final states with two or three soft leptons and missing transverse momentum in proton-proton collisions at $\sqrt{s} =$ 13 TeV}",
  journal        = "JHEP",
  volume         = "04",
  year           = "2022",
  pages          = "091",
  doi            = "10.1007/JHEP04(2022)091",
  eprint         = "2111.06296",
  archivePrefix  = "arXiv",
  primaryClass   = "hep-ex"
}

@article{Duan:2018rls,
  author         = "Duan, Guo-Hua and Hikasa, Ken-ichi and Ren, Jian and Wu, Lei and Yang, Jin Min",
  title          = "{Probing bino-wino coannihilation dark matter below the neutrino floor at the LHC}",
  journal        = "Phys. Rev. D",
  volume         = "98",
  year           = "2018",
  number         = "1",
  pages          = "015010",
  doi            = "10.1103/PhysRevD.98.015010",
  eprint         = "1804.05238",
  archivePrefix  = "arXiv",
  primaryClass   = "hep-ph"
}
\end{document}